\documentclass[twocolumn]{autart}    % Enable this line and disable the 
\usepackage{amssymb,amsmath,amsfonts}
\usepackage{algorithm,algorithmic}
\usepackage{cite}
\usepackage{float}
\usepackage{epsfig}
\usepackage{graphicx}
\usepackage{mathtools}

\usepackage{graphics}
\usepackage{multicol}
\usepackage{lipsum}
\usepackage{subfigure}
\usepackage{url}
\usepackage{verbatim}
\usepackage{xspace}
\usepackage[usenames,dvipsnames]{color}
\usepackage{setspace}
\usepackage{dblfloatfix}

\usepackage{multirow} 

\usepackage{tikz} 
\usetikzlibrary{arrows,quotes}

\floatstyle{ruled}
\newfloat{model}{H}{mod}
\floatname{model}{\footnotesize Model}
\newfloat{notatio}{H}{not}
\floatname{notatio}{\footnotesize Notation}

\newenvironment{varalgorithm}[1]
  {\algorithm\renewcommand{\thealgorithm}{#1}}
  {\endalgorithm}

\newenvironment{list4}{
	\begin{list}{$\bullet$}{%
			\setlength{\itemsep}{0.05cm}
			\setlength{\labelsep}{0.2cm}
			\setlength{\labelwidth}{0.3cm}
			\setlength{\parsep}{0in} 
			\setlength{\parskip}{0in}
			\setlength{\topsep}{0in} 
			\setlength{\partopsep}{0in}
			\setlength{\leftmargin}{0.16in}}}
	{\end{list}}

\newenvironment{list4a}{
	\begin{list}{$\bullet$}{%
			\setlength{\itemsep}{0.05cm}
			\setlength{\labelsep}{0.2cm}
			\setlength{\labelwidth}{0.3cm}
			\setlength{\parsep}{0in} 
			\setlength{\parskip}{0in}
			\setlength{\topsep}{0in} 
			\setlength{\partopsep}{0in}
			\setlength{\leftmargin}{0.16in}}}
	{\end{list}}

\usepackage{url,changebar,bm,xspace,dsfont}
\let\mathbb=\mathds % I much prefer the dsfont over the bbfont
\newtheorem{theorem}{Theorem}

\newtheorem{remark}{Remark}

\usepackage{graphics} % for pdf, bitmapped graphics files
\usepackage{epsfig} % for postscript graphics files
\usepackage{algorithm,algorithmic}

\usepackage{amsmath,amsfonts,amssymb,amscd}
\usepackage{capt-of}

\usepackage{color}
\usepackage{cite}
\usepackage{verbatim}
\newcommand{\AR}[1]{\textcolor{blue}{{#1}}}

\newtheorem{assumption}{\bfseries Assumption}

\begin{document}

\begin{frontmatter}
%\runtitle{Insert a suggested running title}  % Running title for regular 
                                              % papers but only if the title  
                                              % is over 5 words. Running title 
                                              % is not shown in output.

\title{Distributed Coordination Algorithms with Efficient Communication for Open Multi-Agent Systems with Dynamic Communication Links and Processing Delays: Extended Version\thanksref{footnoteinfo}} % Title, preferably 2 lines not more 

\thanks[footnoteinfo]{Preliminary results of this work were presented at the $2025$ IEEE Conference on Decision and Control \cite{Jiaqi_Johan_Apo_Open}. 
The current version of our paper includes (i) the complete and updated convergence proof of the algorithm in \cite{Jiaqi_Johan_Apo_Open}, (ii) a second distributed coordination algorithm (along with its required conditions for convergence) operating over OMAS with dynamic communication links for the case where nodes exhibit processing delays, (iii) a third distributed coordination algorithm (along with its required topological conditions for convergence) operating over indefinitely open multi-agent systems with dynamic communication links, and (iv) an application for distributed sensor fusion for environmental monitoring. 
\\ Corresponding author: Apostolos~I.~Rikos.
\\ The work of A.I.R. was supported by the Guangdong Provincial Project (Grant No. 2024QN11G109).
} 

% \thanks{\ Corresponding author: Apostolos~I.~Rikos.}
% \thanks{This work was supported by the Knut and Alice Wallenberg Foundation and the Swedish Research Council.}

\author[First]{Jiaqi Hu}\ead{jhu021@connect.hkust-gz.edu.cn}, 
\author[Second]{Karl~H.~Johansson}\ead{kallej@kth.se},  
\author[First,Third]{Apostolos~I.~Rikos}\ead{apostolosr@hkust-gz.edu.cn} 

\address[First]{AI Thrust, Information Hub, The Hong Kong University of Science and Technology (Guangzhou), Guangzhou, China}  % Please supply  
\address[Second]{Division of Decision and Control Systems, KTH Royal Institute of Technology, \\ and also affiliated with Digital Futures, SE-100 44 Stockholm, Sweden} 
\address[Third]{Department of Computer Science and Engineering, The Hong Kong University of Science and Technology, \\ Hong Kong SAR, China}  

\begin{abstract}  
In this paper we focus on the distributed quantized average consensus problem in open multi-agent systems consisting of dynamic directed communication links among active nodes. 
We propose three communication-efficient distributed algorithms designed for different scenarios. 
Our first algorithm solves the quantized averaging problem over the currently active node set under finite network openness (i.e., when the active set eventually stabilizes). 
Our second algorithm extends the aforementioned approach for the case where nodes suffer from arbitrary bounded processing delays. 
Our third algorithm operates over indefinitely open multi-agent networks with dynamic communication links (i.e., with continuous node arrivals and departures), computing the average that incorporates both active and historically active nodes. 
We analyze our algorithms’ operation, establish their correctness, and present novel necessary and sufficient topological conditions ensuring their finite-time convergence. 
Numerical simulations on distributed sensor fusion for environmental monitoring demonstrate fast finite-time convergence and robustness across varying network sizes, departure/arrival rates, and processing delays. 
Finally, it is shown that our proposed algorithms compare favorably to algorithms in the existing literature. 
\end{abstract} 


\end{frontmatter}

% ===============================================
%
%
% INTRODUCTION
%
%
% ===============================================
\section{Introduction}\label{sec:intro}

In recent years, multi-agent systems (MAS) have attracted substantial attention from the scientific community due to rising interest in control and coordination algorithms.  
A MAS comprises interacting nodes that collaborate to achieve a shared objective. 
Examples include social networks, networks of sensors or computing devices, and groups of vehicles or robots \cite{2007:olfati-saber_consensus}.

The distributed average consensus problem is a central topic in distributed control, where each node starts with an initial state, and the goal is to compute the average through information exchange among neighbors (see \cite{2018:BOOK} and references therein). 
This problem has attracted extensive research across various settings, including real-valued communication \cite{SEYBOTH:2013}, quantized communication \cite{2021:Rikos_Hadj_Accumul_TAC}, and unreliable networks \cite{chrisTAC:2016}. 

More recently, research efforts have focused to algorithms for MAS capable of managing more complex settings in which nodes may dynamically join or leave the network. 
This class of systems, is referred to as open multi-agent systems (OMAS) \cite{Vizuete2024}. 
Existing research on OMAS can be broadly divided into methods designed for undirected communication links and methods designed for directed communication among nodes. 
Regarding algorithms on OMAS with undirected communication links, early studies proposed pairwise average gossip frameworks \cite{2017_Hendrickx_Martin_CDC}. 
The behavior of pairwise gossip algorithms and dynamical properties of OMAS were analyzed in \cite{2016_Hendrickx_Allerton} and \cite{2021_Franceschelli_Frasca_TAC}, respectively. 
Further developments include distributed mode computation \cite{2022_Dashti_Mauro_IEEELCSS} and heterogeneous pairwise interaction frameworks \cite{2024_Oliva_Scala_TAC_Open}. 
Graphon-based modeling and analysis were introduced in \cite{Vizuete_Hendrickx_graphons2025}, while \cite{Jia_Zhe2025} established consensus conditions for OMAS. 
Additionally, previous works also focused on distributed optimization \cite{2020_Hendrickx_Rabbat_CDC, 2023_Hayashi_TAC, deplano2025optimization}. 
Since most earlier works focused on undirected communication networks, more recent studies have relaxed this assumption and focus on OMAS with directed communication.
Advances in this direction include running-sum averaging schemes \cite{2024:CDC_Hadjic_Garcia}, feedback-assisted averaging \cite{2024:CDC_Themis_Open}, and distributed optimization with dynamic cluster formation \cite{makridis2025_optim_open_multicluster}. 

% Time delays (i.e., transmission delays and processing delays) in information exchange between agents pose a fundamental challenge in consensus problems. 
% These delays, arising from actuation, control, communication, and computation processes, are inherent to practical distributed systems~\cite{cao2012overview}. 
% Extensive research has addressed average consensus under transmission delays: \cite{bliman2008average} analyzed uniform and time-varying delays in undirected networks; \cite{hadjicostis2013average, charalambous2014average} achieved asymptotic consensus considering unknown bounded delays in directed networks and dynamic directed networks, respectively; \cite{charalambous2015distributed} obtained exact finite-time consensus under bounded delays; \cite{makridis2023utilizing} incorporated error control for unreliable communication links; and \cite{charalambous2019privacy} developed privacy-preserving mechanisms. 
% Extensions to integer weight balancing problems have also been explored~\cite{rikos2015integer, rikos2017distributed, rikos2020distributed}. 
% Despite this progress, existing work focuses exclusively on transmission delays while neglecting processing delays. 

It is important to note that current approaches in the literature are subject to several limitations. 
Primarily, most algorithms necessitate the exchange of real-valued messages. 
This results in significant bandwidth consumption, poor resource efficiency, and limits their applicability in real-world scenarios (e.g., bandwidth constrained networks). 
A second major drawback is the lack of algorithms exhibiting finite-time convergence guarantees. 
This limitation is critical for applications requiring nodes to reach agreement promptly and switch to another task. 
Furthermore, existing studies typically assume reliable connectivity between active nodes overlooking potential link variations due to agent mobility or environmental interference. 
This assumption further limits their applicability in real life scenarios. 
Finally, existing algorithms for OMAS assume instantaneous information processing at each node, overlooking realistic processing delays between receiving information and transmitting updates. 

To the best of our knowledge, the aforementioned challenges remain largely unaddressed in the OMAS literature (with the exception our work in \cite{Jiaqi_Johan_Apo_Open}). 
As a result, the development of communication-efficient, finite-time convergent algorithms for OMAS that are robust to dynamic communication topologies and processing delays constitutes a significant open problem. 

\vspace{-.2cm}

\noindent 
\textbf{Main Contributions.}
In this paper, we aim to overcome the limitations of existing approaches by developing three novel distributed quantized averaging algorithms tailored to OMAS with dynamic directed communication links and processing delays. 
Our main contributions are the following: 
\\ \textbullet \ We propose a distributed quantized averaging algorithm operating over OMAS with dynamic communication links (see algorithm QAOD). 
We present its necessary and sufficient topological conditions for convergence (see Assumption~\ref{strong_connectivity_stable_union_graph}) and establish its finite time convergence to the quantized average of the nodes' initial states (see Theorem~\ref{main_convergence_condition_theorem}). 
\\ \textbullet \ We enhance QAOD to operate under bounded, unknown processing delays (see algorithm QAPOD). 
Under this more realistic scenario, we establish our algorithm's finite-time convergence to the quantized average of the nodes' initial states (Theorem~\ref{main_convergence_condition_theorem_process_delays}).
\\ \textbullet \ We propose a distributed quantized averaging algorithm that operates over indefinitely OMAS with dynamic communication links and computes the average over all currently and historically active nodes (see algorithm QAIOD). 
We derive its necessary and sufficient topological conditions for convergence (Assumption~\ref{open_strong_connectivity_stable_union_graph}) and prove its finite-time convergence to the exact quantized average of the initial states (Theorem~\ref{main_convergence_condition_theorem_alwaysopen}).
\\ \textbullet \ We evaluate the performance of our proposed algorithms in a distributed sensor fusion application for environmental monitoring, analyzing their behavior across various network sizes, arrival/departure rates, and processing delays.

Our proposed algorithms build on the quantized averaging algorithm in~\cite{2022:Rikos_Hadj_Johan} with some modifications, since \cite{2022:Rikos_Hadj_Johan} is developed for closed networks with a fixed node set, and therefore does not address the main challenges that arise in OMAS (where nodes may join and leave and where information can be permanently lost at departures if not handled explicitly). 
Our key innovation lies in designing novel strategies that preserve the relevant global sums despite node arrivals and departures, time-varying directed communication links that need not be strongly connected at every time step, and nonzero processing delays. 
Specifically, for our QAOD algorithm we introduce an arrival initialization rule that injects appropriately scaled mass so that a newly arriving agent can be incorporated without violating global sum-preservation, and a departure handoff rule that transfers the departing agent's accumulated mass to remaining agents so that information is not lost at departure. 
Additionally, we establish novel necessary and sufficient topological conditions that rely on joint strong connectivity over time of the directed communication links among the eventually remaining agents, and explicit information retention requirements induced by openness, (i.e., that departing agents have appropriate remaining out-neighbors to which they can hand off their stored mass so that their contribution is not irreversibly lost). 
For our QAPOD algorithm, we introduce a departing soon and long-term-remaining strategy that restricts each node’s transmissions to out-neighbors that are guaranteed to remain active for at least the maximum processing delay horizon, ensuring that any delayed message is delivered and processed before a potential departure and thus preventing mass loss caused by departures within the delay window. 
Finally, for our QAIOD algorithm, we introduce novel arrival initialization and departure handoff rules so that the preserved global sums correspond to the aggregate over all currently and historically active agents, enabling exact quantized averaging in indefinitely open multi-agent systems with dynamic communication links with continual arrivals and departures. 
In addition, we establish novel necessary and sufficient topological conditions that rely on a local out-neighborhood handoff connectivity requirement for departing agents (namely, that every departing node has at least one out-neighbor that remains active so it can transfer its stored mass), together with a global joint strong connectivity condition (i.e., strong connectivity of the virtual union digraph over recurring topology instances) that ensures information propagation and finite-time convergence under dynamic directed links.

Our proposed QAOD, QAPOD, and QAIOD are able to operate over OMAS that consist of directed communication links. 
Additionally (in contrast with most algorithms in the literature) our algorithms are able to operate over networks with dynamic directed communication links among active nodes, and the underlying network does not need to be strongly connected at every time step. 
These advantages greatly enhance the applicability and robustness of our algorithms as practical coordination mechanisms for networks encountered in real-world applications. 
Furthermore, our proposed algorithms are the first in the literature that enable nodes to exchange quantized-valued messages and still achieve finite-time convergence while operating over OMAS with dynamic directed communication links. 
This significantly improves communication efficiency and makes them attractive for bandwidth- and energy-constrained systems. 
Moreover, QAIOD is the first algorithm that provides finite-time convergence guarantees while explicitly handling indefinite openness of the underlying OMAS with dynamic communication links, a feature that naturally arises in many real-world systems with continual agent arrivals and departures. 
Finally, QAPOD is the first to consider processing delays at each node while operating over an OMAS with dynamic communication links. 
This is a realistic phenomenon in computing and communication platforms (where local computation and queuing times are non-negligible), and incorporating this effect constitutes a significant step forward in distributed algorithms for open multi-agent networks and further broadens their applicability in practice. 

\textbf{Paper Organization.} 
The remainder of this paper is organized as follows. 
In Section~\ref{sec:preliminaries} we introduce the necessary notation and background. 
In Section~\ref{sec:probForm} we present the two problems addressed in this work. 
In Sections~\ref{sec:open_dyn_networks}, \ref{sec:open_dyn_network_process_del}, \ref{sec:always_open_dyn_net} we present our three algorithms along with their necessary and sufficient topological conditions, their correctness proofs, and the convergence analyses. 
In Section~\ref{results} we present an application to demonstrate the operation our algorithms, and we compare their operation against other algorithms from the literature. 
In Section~\ref{future} we conclude the paper with a discussion of future research directions. 

% ===============================================
%
%
% RELATED WORKS
%
%
% ===============================================
\section{RELATED WORKS}\label{sec:related_works} 

In this section, we compare the operation of our proposed algorithms with several existing methods from the literature. 
We then provide a summary table (see Table~\ref{table_compar} below) that offers a clearer overview of these comparisons and highlights the advantages of our algorithms over the existing ones. 

A variety of algorithms have been proposed in the literature, each with distinct characteristics and advantages. 
The work in \cite{huynh2006integrated} introduces the FIRE model, which combines four trust and reputation sources into a unified, decentralized scheme for open multi-agent systems operating over undirected graphs. 
In \cite{2017_Hendrickx_Martin_CDC}, the authors develop a framework for open multi-agent systems with undirected, real-valued communication, where nodes perform pairwise gossip averaging while random arrivals and departures continually modify the network size and composition. 
The work in \cite{2022_Dashti_Mauro_IEEELCSS} proposed a framework for mode (majority) computation in open multi-agent systems where nodes interact over an undirected graph and may join or leave the network. 
The key idea was to encode categorical choices as indicator vectors and execute an average-preserving consensus protocol so that they asymptotically agree on the current mode despite arrivals and departures. 
A general framework for open multi-agent systems on undirected graphs with nonlinear, time-varying, heterogeneous couplings is presented in \cite{2024_Oliva_Scala_TAC_Open}. 
The main idea is a per-neighbor storage state that cancels the effect of past interactions with leaving agents. 
The authors illustrate the approach on average consensus, providing conditions for asymptotic convergence once the network topology stops changing.
In \cite{Vizuete_Hendrickx_graphons2025} the authors use graphons to model open multi-agent systems over undirected graphs whose topology is randomly sampled and whose size changes due to agent arrivals, and departures. 
They derive upper bounds on the expected disagreement for both replacement and arrival/departure processes and show that these bounds can be computed based on a matrix whose dimension depends only on the graphon. 
Consensus for open multi-agent systems with time-varying topologies is analyzed in \cite{Jia_Zhe2025}. 
Explicit conditions on switching frequency and on the maximum time the network can remain disconnected are derived that still guarantee consensus. 
In contrast, our proposed QAOD, QAPOD, and QAIOD algorithms explicitly focus on quantized averaging (i.e., exhibit efficient communication) in OMAS and provide finite-time exact quantized averaging guarantees under directed, dynamically changing communication links (features not addressed by the above undirected/real-valued or asymptotic frameworks). 
Research over OMAS has also extended to distributed optimization. 
Specifically, the work in \cite{2020_Hendrickx_Rabbat_CDC} analyzed decentralized gradient descent in open multi-agent systems where nodes (and thus local objective functions) may be added or removed over time. 
It quantified how much the global minimizer can shift when a function is added/removed and proved that DGD iterates remain uniformly bounded regardless of the arrival/departure sequence. 
A distributed subgradient method for distributed optimization in open multi-agent networks with undirected communication links is presented in \cite{2023_Hayashi_TAC}. 
It is shown that the regret grows sublinearly, and the distributed strategy asymptotically approaches the optimal cumulative cost despite the openness of the system. 
In \cite{2023_Nakamura_Inuiguchi_IEEECSS} the authors propose a cooperative decision making method for an adversarial bandit problem in open multi-agent networks with undirected communication. 
Active agents share their local reward estimates with neighbors to update their arm-selection probabilities, and under a stated connectivity and step-size condition the algorithm achieves a sublinear pseudo-regret while enabling agents to collectively identify the best arm. 
The work in \cite{deplano2025optimization} introduces an open operator framework and an Open ADMM algorithm for distributed optimization in open multi-agent systems with undirected communication among nodes. 
It shows that, the Open ADMM algorithm enables nodes to track the time-varying optimizer within a bounded error. 
Our proposed algorithms instead focus on distributed estimation/consensus (namely, exact quantized averaging) in OMAS, and unlike existing distributed optimization methods which typically assume real-valued exchanges over undirected graphs, our QAOD, QAPOD, and QAIOD use quantized messages and provide finite-time guarantees for this objective over dynamic directed communication among the active agents.  
It is important to note that most prior works considered undirected communication networks. 
However, recent studies have relaxed this assumption and handle OMAS with directed communication.
Specifically, \cite{2021_Franceschelli_Frasca_TAC} introduces a general stability framework for open multi-agent networks with directed communication. 
It models arrivals and departures as disturbances and derives Lyapunov-like sufficient conditions for stability under contractivity assumptions. 
In \cite{2024:CDC_Hadjic_Garcia} the authors proposed a modified running-sum ratio consensus algorithm for open multi-agent systems with directed communication. 
Under connectivity assumptions, they show that once the set of participating agents stabilizes, the remaining agents asymptotically reach exact average consensus on their current values. 
The work in \cite{2024:CDC_Themis_Open} addresses average consensus in open multi-agent systems over strongly connected networks with directed communication. 
It proposes an acknowledgement-based ratio consensus algorithm (OPENRC) that preserves total mass and, under column-stochasticity and connectivity assumptions, ensures convergence of all active agents’ states to the desired average once the set of participating agents stabilizes. 
Finally, distributed optimization in open multi-agent systems with directed communication is studied in \cite{makridis2025_optim_open_multicluster} and in \cite{2025_Sawamura_Inuiguchi}. 
In \cite{makridis2025_optim_open_multicluster} the authors propose a gradient-tracking algorithm that uses acknowledgement messages and max-consensus to handle unknown out-neighbor sets and departures, and show that, once cluster membership stabilizes, agents in each cluster converge to the corresponding cluster-wide minimizer. 
The work in \cite{2025_Sawamura_Inuiguchi} proposes a distributed primal-dual algorithm for constrained convex optimization over OMAS with directed communication links. 
The proposed algorithm achieves sublinear bounds in dynamic regret and the constraint
violation under bounded network fluctuations, with validation via an economic dispatch problem in a power network. 
In contrast, our algorithms operate over OMAS with directed communication that are either indefinitely open (see algorithm QAIOD) or eventually become closed (see algorithms QAOD, QAPOD). 
The underlying network is not required to be strongly connected at every time step, and no restrictive network assumptions are imposed (e.g., weight-balanced network or doubly-stochastic mixing weights). 
Additionally, our algorithms exhibit efficient quantized communication, guarantee finite-time convergence to the exact quantized average, and are the first to handle the case of dynamically changing directed links among active nodes. 
Finally, they are the first OMAS quantized-averaging algorithms that explicitly account for processing delays, preventing mass loss within the delay horizon. 

% In Table~\ref{table_compar} we present a summary of our comparisons. 

\begin{table}[h] 
\centering
\caption{Comparison of our algorithms with other works in the literature. 
The metrics are: (A) utilization of directed network, (B) absence of connected, strongly connected, weight balanced network at every time step (C) finite time convergence, (D) efficient communication, (E) dynamic communication links among active nodes, (F) processing delays, and (G) indefinite openness.}
\label{table_compar} 
\begin{tabular}{|c|c|c|c|c|c|c|c|} 
\hline 
\multirow{2}{*}{} 
& (A)    & (B)    & (C)    & (D)    & (E) & (F) & (G)    \\ \hline
\cite{huynh2006integrated}    & \textcolor{red}{\text{\sffamily X}}    & \textcolor{red}{\text{\sffamily X}}    & \textcolor{red}{\text{\sffamily X}}   & \textcolor{red}{\text{\sffamily X}}    & \textcolor{red}{\text{\sffamily X}}  & \textcolor{red}{\text{\sffamily X}}  & \textcolor{green}{\checkmark}    \\ 
\hline
\cite{2017_Hendrickx_Martin_CDC}    & \textcolor{red}{\text{\sffamily X}}    & \textcolor{red}{\text{\sffamily X}}    & \textcolor{red}{\text{\sffamily X}}   & \textcolor{red}{\text{\sffamily X}}    & \textcolor{red}{\text{\sffamily X}}  & \textcolor{red}{\text{\sffamily X}}  & \textcolor{green}{\checkmark}    \\ 
\hline
\cite{golpayegani2019using}    & \textcolor{red}{\text{\sffamily X}}    & \textcolor{red}{\text{\sffamily X}}    & \textcolor{red}{\text{\sffamily X}}   & \textcolor{red}{\text{\sffamily X}}    & \textcolor{red}{\text{\sffamily X}}  & \textcolor{red}{\text{\sffamily X}}  & \textcolor{green}{\checkmark}    \\ 
\hline
\cite{2016_Hendrickx_Allerton}    & \textcolor{red}{\text{\sffamily X}}    & \textcolor{red}{\text{\sffamily X}}    & \textcolor{red}{\text{\sffamily X}}   & \textcolor{red}{\text{\sffamily X}}    & \textcolor{red}{\text{\sffamily X}}  & \textcolor{red}{\text{\sffamily X}}  & \textcolor{green}{\checkmark}    \\ 
\hline
\cite{2021_Franceschelli_Frasca_TAC}    & \textcolor{green}{\checkmark}    & \textcolor{green}{\checkmark}    & \textcolor{red}{\text{\sffamily X}}   & \textcolor{red}{\text{\sffamily X}}    & \textcolor{red}{\text{\sffamily X}}  & \textcolor{red}{\text{\sffamily X}}  & \textcolor{green}{\checkmark}    \\ 
\hline
\cite{2022_Dashti_Mauro_IEEELCSS}    & \textcolor{red}{\text{\sffamily X}}    & \textcolor{red}{\text{\sffamily X}}    & \textcolor{red}{\text{\sffamily X}}   & \textcolor{red}{\text{\sffamily X}}    & \textcolor{red}{\text{\sffamily X}}  & \textcolor{red}{\text{\sffamily X}}  & \textcolor{red}{\text{\sffamily X}}    \\ 
\hline
\cite{2024_Oliva_Scala_TAC_Open}    & \textcolor{red}{\text{\sffamily X}}    & \textcolor{green}{\checkmark}    & \textcolor{red}{\text{\sffamily X}}   & \textcolor{red}{\text{\sffamily X}}    & \textcolor{red}{\text{\sffamily X}}  & \textcolor{red}{\text{\sffamily X}}  & \textcolor{red}{\text{\sffamily X}}    \\ 
\hline
\cite{Vizuete_Hendrickx_graphons2025}    & \textcolor{red}{\text{\sffamily X}}    & \textcolor{green}{\checkmark}    & \textcolor{red}{\text{\sffamily X}}   & \textcolor{red}{\text{\sffamily X}}    & \textcolor{red}{\text{\sffamily X}}  & \textcolor{red}{\text{\sffamily X}}  & \textcolor{green}{\checkmark}    \\ 
\hline
\cite{Jia_Zhe2025}    & \textcolor{red}{\text{\sffamily X}}    & \textcolor{green}{\checkmark}    & \textcolor{red}{\text{\sffamily X}}   & \textcolor{red}{\text{\sffamily X}}    & \textcolor{red}{\text{\sffamily X}}  & \textcolor{red}{\text{\sffamily X}}  & \textcolor{green}{\checkmark}    \\ 
\hline
\cite{2020_Hendrickx_Rabbat_CDC}    & \textcolor{red}{\text{\sffamily X}}    & \textcolor{red}{\text{\sffamily X}}    & \textcolor{red}{\text{\sffamily X}}   & \textcolor{red}{\text{\sffamily X}}    & \textcolor{red}{\text{\sffamily X}}  & \textcolor{red}{\text{\sffamily X}}  & \textcolor{green}{\checkmark}    \\ 
\hline
\cite{2023_Hayashi_TAC}    & \textcolor{red}{\text{\sffamily X}}    & \textcolor{green}{\checkmark}    & \textcolor{red}{\text{\sffamily X}}   & \textcolor{red}{\text{\sffamily X}}    & \textcolor{red}{\text{\sffamily X}}  & \textcolor{red}{\text{\sffamily X}}  & \textcolor{green}{\checkmark}    \\ 
\hline
\cite{2023_Nakamura_Inuiguchi_IEEECSS}    & \textcolor{red}{\text{\sffamily X}}    & \textcolor{green}{\checkmark}    & \textcolor{red}{\text{\sffamily X}}   & \textcolor{red}{\text{\sffamily X}}    & \textcolor{red}{\text{\sffamily X}}  & \textcolor{red}{\text{\sffamily X}}  & \textcolor{green}{\checkmark}    \\ 
\hline
\cite{deplano2025optimization}    & \textcolor{red}{\text{\sffamily X}}    & \textcolor{red}{\text{\sffamily X}}    & \textcolor{red}{\text{\sffamily X}}   & \textcolor{red}{\text{\sffamily X}}    & \textcolor{red}{\text{\sffamily X}}  & \textcolor{red}{\text{\sffamily X}}  & \textcolor{green}{\checkmark}    \\ 
\hline
\cite{2024:CDC_Hadjic_Garcia}    & \textcolor{green}{\checkmark}    & \textcolor{green}{\checkmark}    & \textcolor{red}{\text{\sffamily X}}   & \textcolor{red}{\text{\sffamily X}}    & \textcolor{red}{\text{\sffamily X}}  & \textcolor{red}{\text{\sffamily X}}  & \textcolor{red}{\text{\sffamily X}}    \\ 
\hline
\cite{2024:CDC_Themis_Open}    & \textcolor{green}{\checkmark}    & \textcolor{red}{\text{\sffamily X}}    & \textcolor{red}{\text{\sffamily X}}   & \textcolor{red}{\text{\sffamily X}}    & \textcolor{red}{\text{\sffamily X}}  & \textcolor{red}{\text{\sffamily X}}  & \textcolor{red}{\text{\sffamily X}}    \\ 
\hline
\cite{makridis2025_optim_open_multicluster}    & \textcolor{green}{\checkmark}    & \textcolor{green}{\checkmark}    & \textcolor{red}{\text{\sffamily X}}   & \textcolor{red}{\text{\sffamily X}}    & \textcolor{red}{\text{\sffamily X}}  & \textcolor{red}{\text{\sffamily X}}  & \textcolor{green}{\checkmark}    \\ 
\hline
\cite{2025_Sawamura_Inuiguchi}    & \textcolor{green}{\checkmark}    & \textcolor{red}{\text{\sffamily X}}    & \textcolor{red}{\text{\sffamily X}}   & \textcolor{red}{\text{\sffamily X}}    & \textcolor{red}{\text{\sffamily X}}  & \textcolor{red}{\text{\sffamily X}}  & \textcolor{green}{\checkmark}    \\ 
\hline
QAOD    & \textcolor{green}{\checkmark}    & \textcolor{green}{\checkmark}    & \textcolor{green}{\checkmark}    & \textcolor{green}{\checkmark}    & \textcolor{green}{\checkmark}  & \textcolor{red}{\text{\sffamily X}}  & \textcolor{red}{\text{\sffamily X}}    \\ 
\hline
QAPOD    & \textcolor{green}{\checkmark}    & \textcolor{green}{\checkmark}    & \textcolor{green}{\checkmark}    & \textcolor{green}{\checkmark}    & \textcolor{green}{\checkmark}  & \textcolor{green}{\checkmark}  & \textcolor{red}{\text{\sffamily X}}    \\ 
\hline
QAIOD    & \textcolor{green}{\checkmark}    & \textcolor{green}{\checkmark}    & \textcolor{green}{\checkmark}    & \textcolor{green}{\checkmark}    & \textcolor{green}{\checkmark}  & \textcolor{red}{\text{\sffamily X}}  & \textcolor{green}{\checkmark}    \\ \hline 
\end{tabular} 
\end{table}

% ===============================================
%
%
% NOTATION
%
%
% ===============================================
\section{NOTATION AND BACKGROUND}\label{sec:preliminaries}

\noindent
\textbf{Notation.}
The sets of real, rational, integer and natural numbers are denoted by $\mathbb{R}$, $\mathbb{Q}$, $\mathbb{Z}$, and $\mathbb{N}$, respectively. 
The symbol $\mathbb{Z}_+$ denotes the set of nonnegative integers. 
% while $\mathbb{Z}_0$ denotes the set of natural numbers that includes zero. 
% \todo{$\mathbb{Z}_+$ and $\mathbb{Z}_0$ are same sets. please use one notation. use only $\mathbb{Z}_+$ and not $\mathbb{Z}_0$ and also delete $\mathbb{Z}_0$ from notation} 
For two sets $A$ and $B$, the symbol $\cap$ denotes set intersection ($A \cap B$ is the set of elements common to both sets $A$ and $B$), the symbol $\cup$ represents set union (where $A \cup B$ is the set containing all elements from both $A$ and $B$), and the symbol $\setminus$ denotes set difference (where $A \setminus B$ is the set of elements that are in $A$ but not in $B$).
For any real number $a \in \mathbb{R}$, the floor $\lfloor a \rfloor$ denotes the greatest integer less than or equal to $a$ while the ceiling $\lceil a \rceil$ denotes the least integer greater than or equal to $a$. 
\vspace{.2cm}

\subsection{OMAS}\label{subsec_open_networks}

In OMAS the communication topology is directed and open. 
Directed topology means the flow of information (or interaction) between nodes follows specific directions (i.e., if node $v_j$ can transmit information to node $v_l$, it does not necessarily imply that node $v_l$ can transmit information to node $v_j$). 
Open topology means that the network allows nodes to have the freedom to enter or leave the network at their discretion. 
OMAS consist of $n$ nodes (where $n \geq 2$) communicating only with their immediate neighbors at any time step $k$. 
The finite set of all $n$ nodes \textit{potentially participating} in the OMAS is captured by $\mathcal{V}^\prime = \{v_1, v_2, . . ., v_n\}$ with its cardinality being $n = |\mathcal{V}^\prime|$. 
However, since nodes enter or leave the network, at each time step $k$ only a subset of nodes $\mathcal{V}^o[k] \subseteq \mathcal{V}^\prime$ is considered active. 
Active nodes at time step $k$ are nodes that are present in the network at time step $k$ and capable of sending or receiving information. 
Conversely, inactive nodes at time step $k$ are nodes that are not present in the network at time step $k$, and are incapable of sending or receiving information. 
We assume that interactions occur only between nodes that are active. 
% \todo{We denote the activation index as $\alpha_j[k] \in \{0, 1\}$. Specifically, $\alpha_j[k] = 1$ if $v_j \in \mathcal{V}^o[k]$ and $\alpha_j[k] = 0$ if $v_j \notin \mathcal{V}^o[k]$ (i.e., $\alpha_j[k] = 1$ if node $v_j$ is active at time step $k$ and $\alpha_j[k] = 0$ if it is inactive).}
Following this, the communication topology of the active nodes is modeled as an open digraph $\mathcal{G}^o_d[k]=(\mathcal{V}^o[k], \mathcal{E}^o[k])$, with $\mathcal{V}^o[k] \subseteq \mathcal{V}^\prime$ denoting the set of active nodes at time step $k$, and $\mathcal{E}^o[k] \subseteq \mathcal{V}^o[k] \times \mathcal{V}^o[k]$ denoting the set of edges between active nodes at time step $k$. 
The cardinality of active nodes at time step $k$ is denoted as $n^o[k] = | \mathcal{V}^o[k] |$. 
A directed edge from node $v_j$ to node $v_l$ is denoted by $m^o_{lj} \triangleq (v_l, v_j) \in \mathcal{E}^o[k]$. 
It captures the fact that node $v_j$ can transmit information to node $v_l$ (but not necessarily the other way around). 
The set of all possible edges in the network is denoted as $\mathcal{E}^o = \cup_{k=0}^{\infty} \mathcal{E}^o[k]$. 
The subset of nodes that can directly transmit information to node $v_j$ is called the set of in-neighbors of $v_j$ and is denoted as $\mathcal{N}^{-,o}_j[k] = \{v_i \in \mathcal{V}^o[k] | (v_j, v_i) \in \mathcal{E}^o[k]\}$. 
The cardinality of $\mathcal{N}^{-,o}_j[k]$ is called the in-degree of $v_j$ denoted as $D^{-,o}_j[k] = | \mathcal{N}^{-,o}_j[k] |$. 
The subset of nodes that can directly receive information from node $v_j$ is called the set of out-neighbors of $v_j$ and is denoted as $\mathcal{N}^{+,o}_j[k] = \{v_l \in \mathcal{V}^o[k] | (v_l, v_j) \in \mathcal{E}^o[k]\}$. 
The cardinality of $\mathcal{N}^{+,o}_j[k]$ is called the out-degree of $v_j$ denoted as $D^{+,o}_j[k] = | \mathcal{N}^{+,o}_j[k] |$. 

At each time step $k$, every node $v_j$ belongs in one of the following three subsets (i.e., it operates according to one of the following three operating modes). 

\textbf{Remaining.} 
This subset is denoted by $\mathcal{R}^o[k]$. 
It comprises nodes that are active at both time steps $k$ and $k+1$. Specifically, $\mathcal{R}^o[k]$ represents the nodes that maintain their active status for time steps $k$ and $k+1$ (i.e., nodes that are active in the network at time step $k$, and are not departing in the next time step $k+1$). 
It is defined as 
\begin{equation}\label{remain_set_defn} 
    \mathcal{R}^o[k] = \mathcal{V}^o[k] \cap \mathcal{V}^o[k+1]. 
\end{equation}
\textbf{Arriving.} 
This subset is denoted by $\mathcal{A}^o[k]$. 
It comprises nodes that transition from inactive to active state at time step $k$. 
Specifically, $\mathcal{A}^o[k]$ consists of nodes that are inactive at time step $k$ but become active at time step $k+1$.
It is defined as 
\begin{equation}\label{arrive_set_defn}
    \mathcal{A}^o[k] = \mathcal{V}^o[k+1] \setminus \mathcal{V}^o[k]. 
\end{equation}
\textbf{Departing.} 
This subset is denoted by $\mathcal{D}^o[k]$. 
It comprises nodes that are active at time step $k$ but become inactive at time step $k+1$. 
Specifically, $\mathcal{D}^o[k]$ represents the nodes that exit the network at time step $k$. 
It is defined as 
\begin{equation}\label{depart_set_defn} 
\mathcal{D}^o[k] = \mathcal{V}^o[k] \setminus \mathcal{V}^o[k+1]. 
\end{equation}
Nodes in an OMAS operate according to the previously described modes, transitioning between states as follows. 
At each time step $k$, the inactive nodes can choose the arriving mode to become active. 
The active nodes can choose (i) the departing mode to leave the network and become inactive, or (ii) the remaining mode to remain active. 

\subsection{Dynamic Networks}\label{subsec_dynam_networks}

For introducing dynamic networks we borrow the following description from  \cite[Section~$2.1$]{2022:Rikos_Hadj_Johan}. 
In dynamic networks the communication topology is directed and dynamic (i.e., communication links among nodes change over time). 
The dynamically changing directed network can be captured by a sequence of directed graphs (digraphs), defined as $\mathcal{G}_d^d[k] = (\mathcal{V}^\prime, \mathcal{E}^d[k])$ (for $k=0, 1, ...$), where $\mathcal{V}^\prime =  \{v_1, v_2, \dots, v_n\}$ is the set of nodes and $\mathcal{E}^d[k] \subseteq \mathcal{V}^\prime \times \mathcal{V}^\prime$ is the set of edges at time step $k$. 
% Note here that the definitions of (i) the cardinality of nodes denoted as $n$, (ii) a directed edge from node $v_i$ to node $v_j$ denoted as $m_{ji}$, (iii) the set of in-neighbors and out-neighbors of a node $v_j$ denoted as $\mathcal{N}^{-,d}_j[k]$ and $\mathcal{N}^{+,d}_j[k]$, and (iv) the cardinalities of $\mathcal{N}^{-,d}_j[k]$ and $\mathcal{N}^{+,d}_j[k]$ denoted as $D^{-,d}_j[k]$ and $D^{+,d}_j[k]$, are defined similarly as in Section~\ref{subsec_open_networks}. 
Note that several key definitions remain consistent with those presented in Section~\ref{subsec_open_networks}. 
These include (i) the cardinality of nodes denoted as $n^d = | \mathcal{V}^\prime |$, (ii) a directed edge from node $v_j$ to node $v_l$ represented as $m^d_{lj}$, (iii) the sets of in-neighbors and out-neighbors of a node $v_j$ denoted as $\mathcal{N}^{-,d}_j[k]$ and $\mathcal{N}^{+,d}_j[k]$, respectively, and (iv) the cardinalities of these neighbor sets $\mathcal{N}^{-,d}_j[k]$ and $\mathcal{N}^{+,d}_j[k]$, represented by $D^{-,d}_j[k]$ and $D^{+,d}_j[k]$, respectively. 
Given a dynamic digraph $\mathcal{G}_d^d[k] = (\mathcal{V}^\prime, \mathcal{E}^d[k])$ for $k = 1, 2, ..., m$, where $m \in \mathbb{N}$, its \textit{union digraph} is defined as $\mathcal{G}^{d, \{1, 2, ..., m\}}_d = (\mathcal{V}^\prime, \cup_{k = 1}^{m} \mathcal{E}^d[k])$. 
A dynamic digraph is \textit{jointly strongly connected} over the interval $k = 1, 2, ..., m$ if its corresponding union graph $\mathcal{G}^{d, \{1, 2, ..., m\}}_d$ forms a strongly connected digraph. 
A digraph is strongly connected if for each pair of nodes $v_j, v_l$, (where $v_j \neq v_l$) there exists a directed \textit{path} from $v_j$ to $v_l$. 
This means that in the union graph, for any two distinct nodes $v_j, v_l \in \mathcal{V}^\prime$, there exists a directed \textit{path} from $v_j$ to $v_l$. 

\subsection{OMAS with Dynamic Communication Links}\label{subsec_open_dynam_networks}

In our current work we focus on networks in which at every time step $k$ (i) the flow of information between nodes is directed, (ii) nodes enter or leave the network, and (iii) communication links among active nodes change over time. 
% Note here that a key similarity between \AR{OMAS} with dynamic communication links and the previously discussed OMAS is the separation of nodes in active and inactive. 
% However, the critical difference between these two network types lies in the behavior of communication links among active nodes. 
% In OMAS, these links remain static over time. 
% In contrast, in \AR{OMAS} with dynamic communication links communication links change dynamically even among nodes that are active in the network. 
Specifically, OMAS with dynamic communication links consist of $n$ nodes (where $n \geq 2$), and can be captured by a sequence of digraphs defined as $\mathcal{G}_d[k]=(\mathcal{V}[k], \mathcal{E}[k])$. 
The subset $\mathcal{V}[k] \subseteq \mathcal{V}^\prime$ denotes the set of active nodes at time step $k$. 
Note that the subsets of active and inactive nodes along with the set of all $n$ potentially participating nodes (captured by $\mathcal{V}^\prime = \{v_1, v_2, . . ., v_n\}$ with $n = |\mathcal{V}^\prime|$) were defined in Section~\ref{subsec_open_networks}. 
The subset $\mathcal{E}[k] \subseteq \mathcal{V}[k] \times \mathcal{V}[k]$ denotes the set of edges between active nodes at time step $k$. 
Note here that $\mathcal{V}[k] = \mathcal{V}^{o}[k]$, and $\mathcal{E}[k] = \mathcal{E}^{o}[k] \cap \mathcal{E}^{d}[k]$, during every time step $k$ (where $\mathcal{V}^{o}[k], \mathcal{E}^{o}[k]$ were defined in Section~\ref{subsec_open_networks}, and $ \mathcal{E}^{d}[k]$ was defined in Section~\ref{subsec_dynam_networks}). 
At every time step $k$, the cardinality of active nodes is denoted as $n[k] = | \mathcal{V}[k] |$. 
A directed edge from node $v_j$ to node $v_l$ is denoted by $m_{lj} \triangleq (v_l, v_j) \in \mathcal{E}[k]$, 
and the cardinality of edges at each time step $k$ is denoted as $m[k] = | \mathcal{E}[k] |$. 
The set of all possible edges in the network is denoted as $\mathcal{E} = \cup_{k=0}^{\infty} \mathcal{E}[k]$. 
The set of in-neighbors of node $v_j$ is denoted as $\mathcal{N}^{-}_j[k] = \{v_i \in \mathcal{V}[k] \ | \ (v_j, v_i) \in \mathcal{E}[k]\}$, and the set of out-neighbors as $\mathcal{N}^{+}_j[k] = \{v_l \in \mathcal{V}[k] \ | \ (v_l, v_j) \in \mathcal{E}[k]\}$. 
Note that $\mathcal{N}^{-}_j[k] = \mathcal{N}^{-,o}_j[k] \cap \mathcal{N}^{-,d}_j[k]$, and $\mathcal{N}^{+}_j[k] = \mathcal{N}^{+,o}_j[k] \cap \mathcal{N}^{+,d}_j[k]$ (where $\mathcal{N}^{-,o}_j[k]$, $\mathcal{N}^{+,o}_j[k]$ were defined in Section~\ref{subsec_open_networks}, and $\mathcal{N}^{-,d}_j[k]$, $\mathcal{N}^{+,d}_j[k]$ were defined in Section~\ref{subsec_dynam_networks}). 
% The in-degree of $v_j$ is denoted as $D^{-}_j[k] = | \mathcal{N}^{-}_j[k] |$. 
% The out-degree of $v_j$ is denoted as $D^{+}_j[k] = | \mathcal{N}^{+}_j[k] |$. 
The in-degree and out-degree of $v_j$ are denoted as $D^{-}_j[k] = | \mathcal{N}^{-}_j[k] |$, and $D^{+}_j[k] = | \mathcal{N}^{+}_j[k] |$, respectively. 

Since we have $\mathcal{V}[k] = \mathcal{V}^{o}[k]$ at every time step $k$, in OMAS with dynamic communication links nodes are categorized into operational modes similar to those in OMAS with static communication links. 
This categorization was described in Section~\ref{subsec_open_networks}. 
Specifically, at each time step $k$, every node belongs to one of the three following subsets: 
Remaining (denoted as $\mathcal{R}[k] = \mathcal{R}^o[k]$), 
Arriving (denoted as $\mathcal{A}[k] = \mathcal{A}^o[k]$), or 
Departing (denoted as $\mathcal{D}[k] = \mathcal{D}^o[k]$). 
The definitions of the aforementioned three subsets are analogous to equations \eqref{remain_set_defn}, \eqref{arrive_set_defn}, and \eqref{depart_set_defn}, respectively. 
Nodes transition between these three subsets as described in Section~\ref{subsec_open_networks}. 
This represents a key similarity between OMAS with dynamic communication links and the previously discussed OMAS in Section~\ref{subsec_open_networks} (which stems from the separation of nodes in active and inactive). 
However, the main difference of OMAS with dynamic communication links (compared to the OMAS in Section~\ref{subsec_open_networks}) lies in the behavior of communication links. 
The links between active nodes also change over time. 
This introduces additional challenges for analyzing network structure and designing distributed algorithms that operate over it.

\subsection{Modeling of Processing Delays and Feedback Channels}\label{subsec_modeling_unreliable_commu_link}

\textbf{Processing Delays.}
For each time step $k$, each node $v_j$ may require an \textit{a priori} unknown but bounded number of time steps $\tau_{j}[k]$ to process the information received from its in-neighbors. 
We have that for every time step $k$ it holds \( 0 \leq \tau_{j}[k] \leq \bar{\tau}_{j} < \infty \) where \( \bar{\tau}_j \) is the maximum processing delay at node \( v_j \). 
The maximum processing delay across all nodes is defined as
\begin{equation}\label{upper_bound_processingdelay_opendynamnamnam}
\bar{\tau} = \max_{v_j \in \mathcal{V}'} \bar{\tau}_{j} . 
\end{equation} 
Under processing delays, when node \( v_j \) wants to transmit data at time step $k$ to node \( v_l \) it processes the information for \( \tau_{j}[k] \) time steps before transmitting. 
Node $v_l$ receives the information at time step \( k + \tau_{j}[k] \) (provided the link \( (v_l,v_j) \) exists at that time). 
% Specifically, node \( v_j \) at time \( k \) receives
% \[
% \left\{ x_{ji}[k] \;\middle|\; 0 \leq s \leq k,\; s+\tau_{i}[s]=k,\; v_i \in \mathcal{N}_j^{-}[k] \cup \{ v_j \} \right\} ,
% \]
% for the case where we consider open dynamic networks. 
% The link between nodes is only required to exist at the time of transmission. 

% \textbf{Message Delays.} 
% The transmission from node $v_i$ to node $v_j$ at time step $k$ is subject to an \textit{a priori} unknown bounded delay $\tau_{ji}[k]$, where $0 \leq \tau_{ji}[k] \leq \bar{\tau}_{ji} < \infty$. 
% The maximum delay across all links is defined as
% \begin{equation}\label{upper_bound_delays_opendynam}
%     \bar{\tau} = \max_{(v_j, v_i) \in \mathcal{E}} \bar{\tau}_{ji} , 
% \end{equation} 
% for the case where we have an open dynamic network.  
% We assume $\tau_{jj}[k] = 0$ for every active node $v_j$ for every time step $k$ (ensuring that each node always has immediate access to its own state).
% Considering message delays, the information available at node $v_j$ at time $k$ consists of all messages received from its in-neighbors (or itself) that arrive exactly at time $k$. 
% Specifically, for open dynamic networks we have that at time step $k$ node $v_j$ receives 
% \[
% \left\{ x_{ji}[s] \;\middle|\; 0 \leq s \leq k,\, s + \tau_{ji}[s] = k,\, v_i \in \mathcal{N}_j^{-}[s] \cup \{ v_j \} \right\} . 
% \]
% Similarly, when $v_j$ transmits a message to node $v_l$, the message experiences a delay of $\tau_{lj}[s]$ before $v_l$ receives it. 

\textbf{Modeling Feedback Channels.}
% \AR{In many communication protocols (such as Transmission Control Protocol (TCP) and High-Level Data Link Control (HDLC)), reliable communication over unreliable links is ensured by the \emph{Automatic Repeat reQuest} (ARQ) mechanism \cite{Krouk:2011, Garcia_Widjaja:2000}. 
% ARQ is an error-control protocol that uses acknowledgments to confirm successful packet delivery. 
% ARQ relies on a lightweight feedback signal sent from the receiver to the transmitter to indicate whether a packet has been correctly received (i.e., it is an acknowledgment signal). 
% This feedback is transmitted over a narrowband channel, carries only a single bit per acknowledgment, and is therefore well suited for low-rate and low-power control signaling. 
% This widely adopted and practical design provides us a means of implementing the feedback channel assumption required below (thus, motivating the following assumption in our network model).}
\AR{Many existing communication protocols, such as Transmission Control Protocol (TCP) and High-Level Data Link Control (HDLC), already ensure reliable communication over unreliable physical links via the \emph{Automatic Repeat reQuest} (ARQ) mechanism \cite{Krouk:2011, Garcia_Widjaja:2000}. 
ARQ is a well-established error-control protocol that relies on a lightweight feedback signal, sent from the receiver back to the transmitter, to acknowledge whether a packet has been correctly received. 
This acknowledgment signal is transmitted over a narrowband channel and carries only a single bit per transmission, making it inexpensive to implement and well suited for low-rate, low-power control signaling in practical deployments. 
Motivated by this widely adopted and practical design, we impose the following assumption in our network model, which can therefore be realized directly using standard, off-the-shelf ARQ-based feedback signaling.}

\begin{assumption}\label{assum_feedback_channel}
For each directed link $(v_l, v_j)$ in the network (i.e., directed link from node $v_j$ to node $v_l$), there exists an error-free narrowband feedback channel from node $v_l$ back to node $v_j$. 
\end{assumption}

\section{Problem Formulation}\label{sec:probForm}

Let us consider an OMAS with dynamic communication links $\mathcal{G}_d[k] = (\mathcal{V}[k], \mathcal{E}[k])$. 
Let us assume that each node \( v_j \in \mathcal{V}^\prime\) is initialized with a quantized state \( x_j \) (for simplicity, we let \( x_j \in \mathbb{Z} \)). 
In this paper, we aim to develop distributed algorithms that allow nodes to address the problems \textbf{P1} and \textbf{P2}  presented below. 
Additionally, for enhancing operational and resource efficiency, our algorithms enable nodes to communicate by exchanging quantized valued messages. 

\textbf{P1.} 
Let \( q[k] \) denote the real-valued average of the initial states of all active nodes at time step \( k \), given by
\begin{equation}\label{goal}
    q[k] = \frac{1}{n[k]} \sum_{v_j \in \mathcal{V}[k]} x_j,
\end{equation}
where \( n[k] \) represents the number of active nodes and \( \mathcal{V}[k] \) is the set of active nodes at time step \( k \). 
We require that there exists \( k_0 \in \mathbb{Z}_+ \) so that every active node \( v_j \in \mathcal{V}[k] \) computes in finite time a quantized value \( q^s_j[k] \) satisfying 
\begin{equation}\label{goal_P1}
q^s_j[k] = \lfloor q[k] \rfloor \quad \text{or} \quad q^s_j[k] = \lceil q[k] \rceil, \quad \forall k \geq k_0 , 
\end{equation}
where $q[k]$ is defined in \eqref{goal}. 

% \textbf{P2.} 
% Let us consider an open dynamic network $\mathcal{G}_d[k] = (\mathcal{V}[k], \mathcal{E}[k])$. 
% Each transmission from node $v_i$ to node $v_j$ at time step $k$ may experience an \textit{a priori} unknown but bounded delay $\tau_{ji}[k]$. 
% Furthermore, acknowledgment signals do not experience delays.  
% We require that there exists \( k_0 \in \mathbb{Z}_+ \) so that every active node \( v_j \in \mathcal{V}[k] \) computes in finite time a quantized value \( q^s_j[k] \) satisfying \eqref{goal_P1}. 

{\bf P2.} 
Let us define $q'[k]$ as the real average of the initial states of all nodes that have been active for at least one time step up to time step \( k \). 
The real average $q'[k]$ is defined as 
\begin{equation}\label{eq:extend_goal} 
    q'[k] = \frac{1}{n'[k]} \sum\limits_{v_j \in \mathcal{H}[k]} x_j,
\end{equation} 
where $\mathcal{H}[k]$ denotes the set of all nodes that have been active for at least one time step up to time \( k \), and is defined as 
\begin{equation}\label{set_active_until_k}
    \mathcal{H}[k] = \bigcup_{t=0}^{k} V[t] , 
\end{equation}
and $n'[k] = |\mathcal{H}[k]|$ denotes the total number of distinct nodes that have been active for at least one time step up to time \( k \). 
Note here that $\mathcal{H}[k] \subseteq \mathcal{V}^\prime$  and \( n'[k] \leq n \), for all \( k \) (let us recall that $n = |\mathcal{V}^\prime|$). 
We require that there exists \( k_0 \in \mathbb{Z}_+ \) so that every active node \( v_j \in \mathcal{V}[k] \) computes in finite time a quantized value \( q^s_j[k] \) satisfying 
\begin{equation}\label{goal_P1_ext}
q^s_j[k] = \lfloor q'[k] \rfloor \quad \text{or} \quad q^s_j[k] = \lceil q'[k] \rceil, \quad \forall k \geq k_0 , 
\end{equation} 
where $q'[k]$ is defined in~\eqref{eq:extend_goal}. 

\AR{The average in \eqref{eq:extend_goal} in problem \textbf{P2} is the historical average over all nodes that have been active at least once.  
Since $\mathcal{H}[k] \subseteq \mathcal{V}'$ and $|\mathcal{V}'| = n < \infty$, the cardinality of the historical set is bounded and it becomes constant after the last first-time arrival. 
Recency-weighted or sliding-window averaging constitutes a different objective and is left for future work.}

% % ===============================================
% %
% %
% % ALGORITHM 1 
% %
% %
% % ===============================================

\section{Quantized Average Consensus in OMAS with Dynamic Communication Links}\label{sec:open_dyn_networks}

In this section we present a distributed algorithm that addresses problem \textbf{P1}. 
Our algorithm is QAOD detailed below. 
Before we present the main functionalities of our algorithm, we impose the following assumptions for the development of our results. 
Together with Assumption~\ref{assum_feedback_channel}, these assumptions remain valid during the implementation of QAOD. 
Any additional assumptions required (or relaxations where certain assumptions do not apply) will be stated explicitly before presenting each specific algorithm in the subsequent sections. 

\begin{assumption}\label{awareness_of_remaining_out_neighbors}
At every time step $k \geq 0$, each active node $v_j \in \mathcal{V}[k]$ knows the set of its remaining out-neighbors $\mathcal{N}^{+}_j[k] \cap \mathcal{R}[k]$ (i.e., out-neighbors that will stay active in the next time step). 
\end{assumption} 

\begin{assumption}\label{existence_stable_time_step} 
There exists a time step $k'$ such that 
\begin{equation}\label{stable_nodepart_eq}
    \mathcal{V}[k] = \mathcal{V}_{\mathcal{R}}, \ \forall k \geq k' , 
\end{equation} 
where $\mathcal{V}_{\mathcal{R}}$ is the set of remaining nodes for time steps $k \geq k'$ (i.e., the subset of active nodes in the OMAS with dynamic communication links stabilizes for time steps $k \geq k'$). 
\end{assumption} 

\AR{\begin{assumption}[$T$-jointly strong connectivity]
\label{strong_connectivity_stable_union_graph}
Let us consider an OMAS with dynamic communication links $\mathcal{G}_d[k] = (\mathcal{V}[k], \mathcal{E}[k])$, with $k'$ and $\mathcal{V}_{\mathcal{R}}$ as defined in Assumption~\ref{existence_stable_time_step}. 
For $k \geq k'$, given the infinite sequence of digraphs $\mathcal{G}_d[k']$, $\mathcal{G}_d[k'+1]$, $\dots$, $\mathcal{G}_d[k]$, $\dots$, describing the dynamic communication topology, there is a finite window length $T \in \mathbb{N}$ and an infinite sequence of time instants $t_0, t_1, \dots, t_m, \dots$, where $t_0 = k'$, such that for any $m \in \mathbb{Z}_+$, we have $1 \leq t_{m+1} - t_m \leq T$ and the union graph $\mathcal{G}_d^{t_m, \dots, t_{m+1}-1} \triangleq (\mathcal{V}_{\mathcal{R}}, \bigcup_{k=t_m}^{t_{m+1}-1} \mathcal{E}[k])$ is equal to the nominal digraph $\mathcal{G}_d = (\mathcal{V}_{\mathcal{R}}, \mathcal{E}^{\mathrm{nom}})$ which is assumed to be strongly connected. 
The diameter of the strongly connected union graph $\mathcal{G}_d^{t_m, \dots, t_{m+1}-1}$ is denoted as $D^{un}$ and is the longest shortest path between any two nodes $v_j, v_i \in \mathcal{V}_{\mathcal{R}}$ (note that $D^{un}$ is also the diameter of the nominal digraph $\mathcal{G}_d$).
\end{assumption}}

\begin{assumption}\label{no_delays_assum}
For every time step $k$, the processing performed by node $v_j$ experiences no delay (i.e., for every node $v_j$ it holds $\tau_{j}[k] = 0$ during every time step $k$). 
\end{assumption}

\AR{Assumption~\ref{awareness_of_remaining_out_neighbors} implies that each active node $v_j \in \mathcal{V}[k]$ knows the set of its remaining out-neighbors $\mathcal{N}^{+}_j[k] \cap \mathcal{R}[k]$, at every time step $k \geq 0$. 
This knowledge is important for guaranteeing that each node transmits toward a node that is active and participating in the operation of the algorithm, rather than toward an inactive node, and, crucially, that a departing node (i.e., one becoming inactive at the next time step) can safely hand off its stored information to an active out-neighbor before leaving, so that this information is not lost. 
Although knowledge of $\mathcal{N}^{+}_j[k] \cap \mathcal{R}[k]$ may appear demanding in a network with directed communication, in practice it admits several lightweight and readily implementable realizations. 
One practical approach relies on a ``distress signal'' (i.e., a special tone transmitted in a dedicated control slot or separate channel at higher power than normal communications, enabling it to reach nearby transmitters \cite{2000:bambos_channel}). 
Nodes can thereby learn their out-degree through simple periodic checks, requiring no additional infrastructure. 
A second, practical approach uses the acknowledgment messages already standard in protocols such as TCP, ARQ/HARQ, and ALOHA (see \cite{2024:CDC_Themis_Open}), transmitted over the same dedicated error-free narrowband feedback channel introduced in Assumption~\ref{assum_feedback_channel}. 
These acknowledgment messages simultaneously confirm successful receipt of information and expose channel errors, and, being narrowband signals, they coexist with the main data channel without introducing interference. 
In this sense, Assumption~\ref{awareness_of_remaining_out_neighbors} can be realized using standard, low-overhead signaling mechanisms already deployed in existing communication systems.
\AR{Regarding communication overhead, the required auxiliary information is local. 
At each time step, an active node only needs the availability/mode status of its  out-neighbors (remaining versus departing for QAOD and QAIOD, and long-term  remaining versus non-long-term-remaining for QAPOD). 
This can be conveyed by a  constant-size control message per relevant directed link (for example, one status  bit, or an ACK augmented with a mode flag). 
Hence, the communication overhead is $\mathcal{O}(D_j^-[k])$ bits for node $v_j$, in addition to the quantized data messages.
Also let us note that if an acknowledgment/status signal is lost, corrupted, or delayed beyond the decision time, the mass-preservation invariant used in our proofs may fail, and the exact finite-time guarantees of our theorems do not apply.}}

Assumption~\ref{existence_stable_time_step} enforces finite openness (i.e., there exists $k'$ after which the active node set remains constant). 
This assumption is primarily imposed to make problem \textbf{P1} well-posed as a finite-time convergence task. 
Note that if $\mathcal{V}[k]$ keeps changing indefinitely, then the desired average also changes and the goal becomes tracking a time-varying target rather than converging to a fixed quantized value. 
This scenario is commonly encountered in the analysis of OMAS, since they are often assumed to eventually become closed~\cite{2024_Oliva_Scala_TAC_Open, 2024:CDC_Themis_Open, 2022_Dashti_Mauro_IEEELCSS, 2024:CDC_Hadjic_Garcia}.  
Additionally, finite openness naturally arises in numerous practical applications, including the following examples: (i) sensor networks conducting periodic monitoring campaigns, where nodes join during initialization, stabilize for data collection and fusion, then may later be shut down, and (ii) wireless edge-computing clusters performing distributed inference, where worker nodes may be added or removed during maintenance or scaling, but consensus computation occurs over a stable set of available resources. 
In all these scenarios, the network temporarily exhibits arrivals and departures, then closes for the duration of the critical consensus phase. 
For applications where openness persists indefinitely (such as social networks with continuous membership changes) we explicitly address this via problem \textbf{P2} and algorithm QAIOD, which computes the average over all nodes that have ever been active (as we show later in Section~\ref{sec:always_open_dyn_net}).

% Assumption~\ref{existence_stable_time_step} states that there exists a time step \( k' \) at which the set of active nodes in the open dynamic network stabilizes (no further arrivals or departures occur). 
% Under this assumption, the average of the initial states of the active nodes remains unchanged for all \( k \geq k' \), enabling nodes to compute this value using our proposed algorithm. 
% This scenario is commonly encountered in the analysis of open networks, since they are often assumed to eventually become closed~\cite{deplano2025optimization, 2024_Oliva_Scala_TAC_Open, 2024:CDC_Themis_Open, 2022_Dashti_Mauro_IEEELCSS, 2024:CDC_Hadjic_Garcia}. 
% However, our algorithm can also be adapted to situations where the network remains open indefinitely (as we will show later in Section~\ref{sec:always_open_dyn_net}). 

Assumption~\ref{strong_connectivity_stable_union_graph} guarantees that, for all time steps \( k \geq k' \), there exists at least one directed path between every pair of nodes infinitely often. 
\AR{In practice, this does not require the network to be strongly connected at every time step, as instantaneous communication networks may be disconnected or sparse due to wireless fading, directional antennas, or mobile nodes temporarily moving out of range. 
Instead, it only requires that over each recurrent window of length at most $T$, the union graph coincides with a fixed strongly connected nominal digraph (which is a widely adopted model for time-varying topologies in the distributed coordination and networked control literature \cite{Nedic:2015TAC}). 
This condition is naturally satisfied without centralized verification, e.g., by mobile agents following patrol trajectories, randomized coverage walks, or multi-hop routing protocols that guarantee recurrent connectivity.  
The window length $T$, together with the diameter $D_{\mathrm{un}}$ of the nominal digraph, jointly determine the finite-time horizon within which information propagates from any node to every other node in $\mathcal{V}_{\mathcal{R}}$, enabling us to establish our finite time convergence guarantees.} 

Assumption~\ref{no_delays_assum} implies that all local computations at each node are performed instantaneously throughout the algorithm's execution (i.e., each node is able to instantaneously process the information received from its in-neighbors). 
% This ensures every node always has immediate access to the latest computed values for transmission. 
This enables our analysis to focus solely on the effects of network topology and node arrivals or departures, without complications arising from processing delays.

\subsection{Distributed Quantized Averaging Algorithm in OMAS with Dynamic Communication Links}\label{subsec_distr_alg_open_dynamic}

Let us consider an OMAS with dynamic communication links $\mathcal{G}_d[k] = (\mathcal{V}[k], \mathcal{E}[k])$. 
Each node $v_j \in \mathcal{V}^\prime$ has an initial quantized state $x_j \in \mathbb{Z}$. 
Additionally, each node $v_j$ maintains the mass variables $y_j[k]$, $z_j[k]$, the state variables $y_j^s[k]$, $z_j^s[k]$, $q_j^s[k]$, and the transmission variables $c_{lj}^y[k]$, $c_{lj}^z[k]$ at each time step $k$. 
The mass variables are utilized to perform computations on the stored information, the state variables are utilized to store the received information and calculate \eqref{goal}, and the transmission variables are utilized to transmit messages to other nodes.  
During the execution of our algorithm, each node executes the following operations. 

\textbf{Remaining Strategy.} 
During every time step $k$, each active node with remaining status $v_j \in \mathcal{R}[k]$ assigns to each of its outgoing edges $m_{lj}$ (including a virtual self-edge) a nonzero probability  
\begin{equation}\label{eq:remain_trans_prob}
     b_{lj}[k] \hspace{-.1cm} = \hspace{-.1cm} \left\{ 
\begin{aligned} 
\frac{1}{1 + |\mathcal{N}^{+}_j[k] \cap \mathcal{R}[k]|} , \ \hspace{-.05cm} & v_l \in (\mathcal{N}^{+}_j[k] \cap \mathcal{R}[k]) \cup \{v_j\}, \\
0, & \ \text{otherwise}.  
\end{aligned}
\right.
\end{equation}
Then, it updates its state variables to be equal to the mass variables and also updates its transmission variables. 
For updating its transmission variables $c_{lj}^y[k]$, $c_{lj}^z[k]$, it splits $y_j[k]$ into $z_j[k]$ equal pieces, keeps one piece for itself and transmits the other pieces to randomly chosen out-neighbors or itself according to the assigned nonzero probabilities $b_{lj}[k]$ above. 
Then, $v_j$ receives the transmission variables from every in-neighbor $v_i \in \mathcal{N}^{-}_j[k]$, and updates its mass variables $y_j[k+1]$, $z_j[k+1]$ as follows:
\begin{equation}\label{eq:mass_var_update}
\begin{aligned}
    y_j[k+1] = & c_{jj}^y[k] + \sum\limits_{v_i \in \mathcal{N}^{-}_j[k]} w_{ji}[k] \ c_{ji}^y[k] \\
    z_j[k+1] = & c_{jj}^z[k] + \sum\limits_{v_i \in \mathcal{N}^{-}_j[k]} w_{ji}[k] \ c_{ji}^z[k]
    \end{aligned}
\end{equation}
where $w_{ji}=1$ if node $v_j$ receives $c_{ji}^y[k]$, $c_{ji}^z[k]$ from $v_i \in \mathcal{N}^{-}_j[k]$ at iteration $k$ (otherwise $w_{ji}[k] = 0$). 
Note that a more detailed description of the operation performed by the remaining nodes $v_j \in \mathcal{R}[k]$ is shown in \cite[Algorithm~$1$]{2022:Rikos_Hadj_Johan}.

\textbf{Arriving Strategy.} 
When node $v_j$ arrives in the network at time step $k$ (i.e., $v_j \in \mathcal{A}[k]$), it updates its state $y_j^s[k+1]$, $z_j^s[k+1]$, $q_j^s[k+1]$, and mass variables $y_j[k+1]$, $z_j[k+1]$, as follows:
\begin{equation}\label{init_variables}
\begin{aligned}
    y_j[k+1] &= 2x_j, \ z_j[k+1] = 2r_j, \\
    y_j^s[k+1] &= 2x_j, \ z_j^s[k+1] = 2r_j, \\
    q_j^s[k+1] &= \left\lfloor \frac{y_j^s[k+1]}{z_j^s[k+1]} \right\rfloor ,  
\end{aligned}
\end{equation}
where $r_j = 1$. 
Then, the node $v_j$ starts interacting with its active neighbors at the next time step. 

\textbf{Departing Strategy.} 
When node $v_j$ departs from the network at time step $k$ (i.e., $v_j \in \mathcal{D}[k]$), it assigns to each of its outgoing edges $m_{lj}$ where $v_l \in \mathcal{N}^{+}_j[k] \cap \mathcal{R}[k]$, a nonzero probability value $b_{lj}[k]$ as follows: 
\begin{equation}\label{eq:depart_trans_prob}
     b_{lj}[k]=\left\{ 
\begin{aligned} 
\frac{1}{|\mathcal{N}^{+}_j[k] \cap \mathcal{R}[k]|} , & \ v_l \in \mathcal{N}^{+}_j[k] \cap \mathcal{R}[k], \\
0, & \ v_l \notin \mathcal{N}^{+}_j[k] \cap \mathcal{R}[k]. 
\end{aligned}
\right.
\end{equation}
Then it randomly selects a \textit{remaining} out-neighbor $v_l \in \mathcal{N}^{+}_j[k] \cap \mathcal{R}[k]$ according $b_{lj}[k]$ and transmits towards $v_l$ its negative two times initial state $-2x_j$ and $-2r_j$ combined with its mass variables $y_j[k]$ and $z_j[k]$. 
Specifically, the computation of transmission variables $c_{lj}^y[k]$, $c_{lj}^z[k]$ is as follows:
\begin{equation}\label{trans_vari}
\begin{aligned}
    &c_{lj}^y[k] = y_j[k] - 2x_j,\\
    &c_{lj}^z[k] = z_j[k] - 2r_j. 
\end{aligned}
\end{equation}

\AR{Note here that at each time step $k$, the local computation of node $v_j$ is $O(D_j^+[k] + D_j^-[k] + z_j[k])$, where $z_j[k] \le 2n$, and it only involves elementary integer operations. 
Additionally, each node stores a constant number of scalar variables, i.e., $O(1)$ memory (or $O(D_j^+[k])$ including the transmission variables). 
Moreover, a remaining node $v_j$ transmits one message, consisting of two quantized integers, to each of its remaining out-neighbors, i.e., $O(D_j^+[k])$ messages per step. 
The total communication until convergence additionally depends on the convergence time, which depends on the network topology and the initial states, and is left for future work.}

In QAOD below we provide a summary of a single step of our proposed procedure. 
This summary outlines all operational modes for each node $v_j$. 

\begin{varalgorithm}{1}
\caption{QAOD} 
\noindent \textbf{Input.} 
An OMAS with dynamic communication links $\mathcal{G}_d[k]=(\mathcal{V}[k], \mathcal{E}[k])$ with $n = | \mathcal{V}^\prime |$ nodes potentially participating, $n[k] = | \mathcal{V}[k] |$ active nodes, and $m[k] = | \mathcal{E}[k] |$ edges at each time step $k$. 
Each potentially participating node $v_j \in \mathcal{V}^\prime$ has initial state $x_j \in \mathbb{Z}$, and $r_j = 1$. 
Also, Assumptions~\ref{assum_feedback_channel}, \ref{awareness_of_remaining_out_neighbors}, \ref{existence_stable_time_step}, \ref{strong_connectivity_stable_union_graph}, \ref{no_delays_assum} hold. 
\\
\textbf{Initialization.} 
Each node $v_j \in \mathcal{V}^\prime$ sets $y_j[0] = 2x_j$, $z_j[0] = 2r_j$. \\ 
%Sets $y_j[0] := \pi^{\mathrm{upper}} (l_j + u_j)$, $z_j[0] = \pi_j^{\max}$, and $\text{flag}_j = 0$. 
\textbf{Iteration.} For each time step $k= 0, 1, 2, \dots$\\
\textbf{Arriving:} Each node $v_j \in \mathcal{A}[k]$ updates its state $y_j^s[k+1]$, $z_j^s[k+1]$, $q_j^s[k+1]$, and mass variables $y_j[k+1]$, $z_j[k+1]$ as in \eqref{init_variables}. \\
\textbf{Departing:} Each node $v_j \in \mathcal{D}[k]$ does:
\begin{list4}
\item[$1)$] Assigns a nonzero probability $b_{lj}[k]$ to each of its outgoing edges $m_{lj}$ 
% where $v_l \in \mathcal{N}^{+}_j[k] \cap \mathcal{R}[k]$ 
as in \eqref{eq:depart_trans_prob}. 
\item[$2)$] Chooses randomly one remaining out-neighbor $v_l \in \mathcal{N}^{+}_j[k] \cap \mathcal{R}[k]$ according to probability $b_{lj}[k]$. 
\item[$3)$] Computes transmission variables $c_{lj}^y[k]$, $c_{lj}^z[k]$ as in \eqref{trans_vari}. 
\item[$4)$] Transmits $c_{lj}^y[k]$, $c_{lj}^z[k]$ to the selected remaining out-neighbor $v_l$. 
\end{list4} 
\textbf{Remaining:} Each node $v_j \in \mathcal{R}[k]$ does:
\begin{list4} 
\item[$1)$] Assigns to each of its outgoing edges $m_{lj}$ (including a virtual self-edge) a nonzero probability $b_{lj}[k]$ as in $\eqref{eq:remain_trans_prob}$. 
\item[$2)$] Sets $y_j^s[k] = y_j[k], z_j^s[k] = z_j[k], q_j^s[k] = \lfloor\frac{y_j^s[k]}{z_j^s[k]}\rfloor$. 
\item[$3)$] Sets $c_{lj}^y[k] = 0$, $c_{lj}^z[k] = 0$, for every $v_l \in (\mathcal{N}^{+}_j[k] \cap \mathcal{R}[k]) \cup \{v_j\}$. 
\item[$4)$] Sets $\delta_z = z_j[k]$. 
\item[$5)$] \textbf{If} $\delta_z \leq 1$ \textbf{set} $c_{jj}^y[k] = y_j[k]$, $c_{jj}^z[k] = z_j[k]$. 
\item[$6)$] \textbf{While} $\delta_z > 1$ \textbf{do} 
\begin{list4a}
\item[$6.1)$] $\delta_y = \lfloor y[k] / z[k] \rfloor$.
\item[$6.2)$] Chooses $v_l \in (\mathcal{N}^{+}_j[k] \cap \mathcal{R}[k]) \cup \{v_j\}$ according to $b_{lj}[k]$.  
\item[$6.3)$] Sets $c_{lj}^y[k] = c_{lj}^y[k] + \delta_y$, and $c_{lj}^z[k] = c_{lj}^z[k] + 1$ for chosen $v_l$ in previous step.   
\item[$6.4)$] Sets $y_j[k] = y_j[k] - \delta_y$, $z_j[k] = z_j[k] - 1$, $\delta_z = \delta_z - 1$. 
\end{list4a}
\item[$7)$] Transmits $c_{lj}^y[k]$ and $c_{lj}^z[k]$, to $v_l$ for every $v_l \in (\mathcal{N}^{+}_j[k] \cap \mathcal{R}[k]) \cup \{v_j\}$. 
\item[$8)$] Receives $c_{lj}^y[k]$, $c_{lj}^z[k]$ from in-neighbor $v_i \in \mathcal{N}^{-}_j[k]$ and updates $y_j[k+1]$, $z_j[k+1]$ as in \eqref{eq:mass_var_update}. 
\end{list4}
\textbf{Output:} Every active node $v_j \in \mathcal{V}[k]$ addresses \textbf{P1}. 
\label{algorithm1} 
\end{varalgorithm}

\textbf{Intuition.} 
The operation of QAOD can be viewed as a ``random walk'' of \( n[k] \) ``tokens'' on a Markov chain of \( n[k] = |\mathcal{V}[k]| \) states. 
In this Markov chain the interconnections (and thus the transition probabilities) vary over time. 
Also, for \( k < k' \), the number of states changes before stabilizing for \( k \geq k' \) (see Assumption~\ref{existence_stable_time_step}). 
Note here that the core averaging mechanism is analyzed in \cite[Theorem~1]{2022:Rikos_Hadj_Johan}. 
Initially, each active node \( v_j \) holds two tokens: a stationary token \( T_j^{ins} \) and a token \( T_j^{out} \) that performs a random walk. 
Each token contains the value pairs \( (y_j^{ins}[k], z_j^{ins}[k]) \) and \( (y_j^{out}[k], z_j^{out}[k]) \). 
These tokens are initialized with $y_j^{ins}[0] = y_j^{out}[0] = x_j \in \mathbb{Z}$ and $z_j^{ins}[0] = z_j^{out}[0] = r_j = 1$. 
During evert time step $k$, the algorithm preserves the sums \( \sum_{v_j \in \mathcal{V}[k]} (y_j^{ins}[k] + y_j^{out}[k]) = \sum_{v_j \in \mathcal{V}[k]} 2x_j \) and \( \sum_{v_j \in \mathcal{V}[k]} (z_j^{ins}[k] + z_j^{out}[k]) = 2n[k] \). 
% The operation of QAOD at any time step \( k \) can be viewed as a ``random walk'' of \( n[k] \) ``tokens'' on a Markov chain with \( n[k] = |\mathcal{V}[k]| \) states. 
% In this Markov chain the interconnections (and thus its transition probabilities) vary over time. 
% Moreover, for time steps \( k < k' \) the number of states in the Markov chain also changes with time, while for \( k \geq k' \) the state set remains fixed (see Assumption~\ref{existence_stable_time_step}). 
% Note that a comprehensive analysis of the underlying averaging algorithm can be found in \cite[Theorem~$1$]{2022:Rikos_Hadj_Johan}. 
% Initially, each active node $v_j$ holds two ``tokens'': $T_j^{ins}$ (which is stationary) and $T_j^{out}$ (which performs a random walk). 
% At every time step $k$, each token contains a pair of values, namely $y_j^{ins}[k]$, $z_j^{ins}[k]$, and $y_j^{out}[k]$, $z_j^{out}[k]$, respectively, for which it holds that $y_j^{ins}[0] = y_j^{out}[0] = x_j \in \mathbb{Z}$ and $z_j^{ins}[0] = z_j^{out}[0] = r_j = 1$. 
% At each time step $k$, the sum of $y^{ins}_j[k] + y^{out}_j[k]$ of all active nodes is equal to $\sum_{v_j \in \mathcal{V}[k]} 2x_j$, and the sum of $z^{ins}_j[k] + z^{out}_j[k]$ is equal to $2n[k]$. 
The operations of each mode can be interpreted as follows. 
\\ \noindent 
\textit{Remaining.} At each time step $k$, each node $v_j$ keeps the token $T_j^{ins}$ (i.e., it never transmits it) and transmits the token $T_j^{out}$.
When $v_j$ receives one or more tokens $T_i^{out}$ from its in-neighbors, the values $y_i^{out}[k]$ and $y_j^{ins}[k]$ become equal (or differ by at most~$1$). 
Then, $v_j$ transmits each received token $T_i^{out}$ to a randomly selected out-neighbor or itself. 
 \\ \noindent 
\textit{Arriving.} Each arriving node $v_j \in \mathcal{A}[k]$ at time step $k$ holds two ``tokens'' $T^{ins}_j$ (which is stationary) and $T^{out}_j$ (which performs a random walk). 
These tokens contain pairs of values $y^{ins}_j[k+1]$, $z^{ins}_j[k+1]$, and $y^{out}_j[k+1]$, $z^{out}_j[k+1]$, respectively, for which it holds that $y^{ins}_j[k+1] = y^{out}_j[k+1] = x_j$ and $z^{ins}_j[k+1] = z^{out}_j[k+1] = r_j = 1$.  
 \\ \noindent 
\textit{Departing.} 
When a node \( v_j \in \mathcal{D}[k] \) departs from the network, its initial tokens must be removed and the initial tokens of the remaining nodes must be preserved. 
To achieve this, each departing node transmits to one remaining out-neighbor a message that combines two values. 
The first value is the negative sum of its initial token values at time step \( k = 0 \), specifically \( -(y^{out}_j[0] + y^{ins}_j[0]) = -2 x_j \) and \( -(z^{out}_j[0] + z^{ins}_j[0]) = -2 r_j \), thereby eliminating the contribution of its own initial tokens. 
The second is the current sum of its token values, \( y_j[k] \) and \( z_j[k] \). 
By performing this transmission it ensures that its initial tokens are removed from the network and also the initial tokens corresponding to the remaining nodes are not lost. 

% \begin{remark}[Relaxation of Strong Connectivity]
%     Existing algorithms in the literature for OMAS with directed communication links typically require the network to be strongly connected at every time step \( k \geq 0 \) (see for example \cite{2024:CDC_Hadjic_Garcia, 2024:CDC_Themis_Open}). 
%     It is important to note that our QAOD (and also QAPOD and QAIOD presented later) relaxes this requirement.
%     During the operation of QAOD the OMAS with dynamic communication links \( \mathcal{G}_d[k] \) is not required to be strongly connected at all time steps. 
%     Instead, we only require its union digraph \( \mathcal{G}_d^{ \{ 1, 2, ..., T \} } \) to be \( T \)-jointly strongly connected for time steps \( k \geq k' \) (see Assumption~\ref{strong_connectivity_stable_union_graph}), in order to guarantee convergence of our proposed distributed algorithm. 
%     This relaxation significantly widens the scope of our algorithms, making them applicable to more realistic settings where network connectivity may be intermittent or time-varying, as it enables convergence under weaker connectivity conditions. 
% \end{remark}

\subsection{Correctness Analysis of QAOD} 

We now establish the correctness of our QAOD via the following theorem. 
Additionally, in our theorem we also provide a necessary and sufficient condition for nodes executing QAOD in an OMAS with dynamic communication links to solve problem \textbf{P1}. 

\begin{theorem}\label{main_convergence_condition_theorem}
Let us consider an OMAS with dynamic communication links $\mathcal{G}_d[k]=(\mathcal{V}[k], \mathcal{E}[k])$ with $n = | \mathcal{V}^\prime |$ nodes potentially participating, $n[k] = | \mathcal{V}[k] |$ active nodes, and $m[k] = | \mathcal{E}[k] |$ edges at each time step $k$.  
Suppose that Assumptions~\ref{assum_feedback_channel}, \ref{awareness_of_remaining_out_neighbors}, \ref{existence_stable_time_step}, \ref{strong_connectivity_stable_union_graph}, \ref{no_delays_assum} hold. 
Let us also assume that all potentially participating nodes execute QAOD. 
There exists a time step $k_0$ such that for every active node $v_j \in \mathcal{V}[k]$ we have 
$$ 
( q^s_j[k] = \lfloor q[k] \rfloor ) \ \text{or} \ ( q^s_j[k] = \lceil q[k] \rceil ) , 
$$
for $k \geq k_0$, where $q[k]$ is defined in \eqref{goal}, if and only if for every departing node $v_j \in \mathcal{D}[k]$ it holds that 
\begin{equation}\label{condition_for_correctness} 
    | \ \mathcal{N}^{+}_j[k] \cap \mathcal{R}[k] \ | \geq 1 \ , 
\end{equation} 
for every time step $k$. 
\end{theorem}

\textbf{PROOF.} See Appendix~\ref{convergence_Alg1_prob_P1}. \hspace*{\fill} $\square$

% \begin{remark}[Intuition of Theorem~\ref{main_convergence_condition_theorem}]
% Theorem~\ref{main_convergence_condition_theorem} establishes that for QAOD to enable active nodes to solve problem~\textbf{P1}, each departing node must have at least one \textit{remaining} out-neighbor. 
% This requirement is critical for preserving relevant information and also for removing obsolete data from the network. 
% If a departing node lacks remaining out-neighbors its stored information is lost. 
% This may disrupt the consistent operation of QAOD and cause active nodes to converge to a consensus value not equal to the quantized average of the initial states. 
% However, if a departing node is able to transmit to at least one remaining node prior to its departure then its initial state is properly removed and vital information is maintained. 
% This guarantees the correct convergence of QAOD.
% \end{remark}

\begin{remark}[Intuition of Theorem~\ref{main_convergence_condition_theorem}]
Theorem~\ref{main_convergence_condition_theorem} establishes that for QAOD to enable active nodes to solve problem~\textbf{P1}, each departing node must have at least one \textit{remaining} out-neighbor. 
This requirement is critical for preserving relevant information and also for removing obsolete data from the network. 
If a departing node lacks remaining out-neighbors its stored information is lost. 
This may disrupt the consistent operation of QAOD and cause active nodes to converge to a consensus value not equal to the quantized average of the initial states. 
However, if a departing node is able to transmit to at least one remaining node prior to its departure then its initial state is properly removed and vital information is maintained. 
This guarantees the correct convergence of QAOD. 
\AR{The proof of Theorem~\ref{main_convergence_condition_theorem} proceeds in two distinct phases:
\\ \textbullet \ \textit{Transient Open Phase ($k < k'$).}
During the phase where node arrivals and departures actively occur (i.e., for $k < k'$), as long as each departing node transfers its state to at least one remaining out-neighbor (the handoff condition above), the aggregate mass across the active network is strictly preserved at every time step. 
This prevents information loss and ensures obsolete states are properly removed.
\\ \textbullet \ \textit{Post-Stabilization Phase ($k \geq k'$).}
Once the active node set stabilizes to $\mathcal{V}_R$ (see Assumption~\ref{existence_stable_time_step}), the communication network operates over recurrent strongly connected union graphs (see Assumption~\ref{strong_connectivity_stable_union_graph}). 
Under this joint-connectivity property, the preserved mass is guaranteed to mix and propagate across the network within bounded time windows, driving the quantized states of all active nodes to the ceiling or floor of the true average in finite time.} 
\end{remark}

% % ===============================================
% %
% %
% % ALGORITHM 2 
% %
% %
% % ===============================================

\section{Quantized Average Consensus in OMAS with Dynamic Communication Links and with Processing Delays}\label{sec:open_dyn_network_process_del}

% In real-world distributed networks, nodes require a nonzero number of time steps to process stored information received from their in-neighbors and/or the operating environment (thus making our Assumption~\ref{no_delays_assum} infeasible). 
% Processing delays can critically affect both the correctness and convergence of distributed algorithms, since nodes may attempt to perform key operations (such as departing from the open dynamic network) before finishing necessary computations and transmitting their updated information to neighbors. 
% Driven by these practical challenges, we now consider the impact of processing delays (as described in Section~\ref{subsec_modeling_unreliable_commu_link}) within open dynamic networks. 
% We propose a distributed algorithm (detailed below as \AR{QAPOD}) that addresses problem \textbf{P1} while explicitly accounting for the possibility that each node may experience an \textit{a priori} unknown but bounded delay for processing the information received from its in-neighbors at each time step. 
% Note here that for our proposed algorithm, Assumptions~\ref{assum_feedback_channel}, \ref{existence_stable_time_step}, and \ref{strong_connectivity_stable_union_graph} are required, while Assumptions~\ref{awareness_of_remaining_out_neighbors}, and \ref{no_delays_assum} do not apply. 
% Additionally, we also impose the following two assumptions that are essential for our subsequent analysis. 

In real-world distributed networks, nodes require nonzero time to process information received from in-neighbors or the environment (rendering Assumption~\ref{no_delays_assum} infeasible). 
Processing delays can critically affect correctness and convergence when nodes, for example, depart before completing computations and transmitting updates. 
Driven by these challenges, we consider processing delays within OMAS with dynamic communication links. 
We propose QAPOD below for addressing problem \textbf{P1} for the case where each node experiences an \textit{a priori} unknown but bounded processing delay. 
For our algorithm, Assumptions~\ref{assum_feedback_channel}, \ref{existence_stable_time_step}, and \ref{strong_connectivity_stable_union_graph} hold, while Assumptions~\ref{awareness_of_remaining_out_neighbors} and \ref{no_delays_assum} are not required. 
We also introduce two additional assumptions below that are necessary for our subsequent analysis.

\begin{assumption}\label{assum:bounded_message_delay}
At each time step \( k \), each node \( v_j \) requires an \textit{a priori} unknown but bounded number of time steps (denoted as \( \tau_{j}[k] \)) to process the information received from its in-neighbors (where \( 0 \leq \tau_{j}[k] \leq \bar{\tau}_j < \infty \)).  
The maximum processing delay among all nodes is given by \( \bar{\tau} \) as defined in~\eqref{upper_bound_processingdelay_opendynamnamnam}.
\end{assumption}

\begin{assumption}\label{assum:know_when_it_leave} 
Each active node $v_j$ at time step $k$ knows whether it will leave the network within the time interval $\{ k + \rho_j^l[k], ..., k+ \rho_j^u[k]\}$, where $\rho_j^u[k] > \rho_j^l[k] \geq 1$. 
Specifically, at time step $k$ node $v_j$ knows whether there exists $s \in \{ k + \rho_j^l[k] , \ldots , k + \rho_j^u[k]\}$ such that $v_j \in \mathcal{D}[s]$. 
\end{assumption}

Assumption~\ref{assum:bounded_message_delay} implies that local computations at each node incur processing delays upper bounded by \( \bar{\tau} \). 
This means that each node typically requires a nonzero (but finite) amount of time to process information received from its in-neighbors before transmitting updated values.
% As a result, our analysis must explicitly account for these per-node delays alongside network topology changes and node arrivals/departures.

Assumption~\ref{assum:know_when_it_leave} ensures that each active node can reliably notify its in-neighbors that it will depart within the window  $\{ k + \rho_j^l[k], \ldots, k+ \rho_j^u[k]\}$.
This holds for pre-scheduled departures (e.g., fixed activity periods, sleep cycles, planned shutdowns), for anticipated shutdowns via local resource monitoring (battery, memory, task completion), and for lightweight forecasting from on-device telemetry (e.g., remaining energy, consumption trends, duty-cycle predictions) that estimates a power-off window with bounded uncertainty. 
Such proactive notification ensures messages are sent only to nodes guaranteed to remain active for a specified number of steps, preventing message loss.

\subsection{Subsets of Nodes Departing Soon and Long-Term Remaining}\label{depart_soon_mode}

So far we have introduced three subsets of nodes (or operating modes). 
Specifically, considering an OMAS with dynamic communication links every node belongs to one of the following three subsets: Remaining $\mathcal{R}[k]$, Arriving $\mathcal{A}[k]$, or Departing $\mathcal{D}[k]$, at each time step $k$. 
However, to handle processing delays, we introduce a new subset (or operating mode) called ``Departing Soon.'' 
This subset is defined below as $\mathcal{S}[k]$ for any time step $k$. 
Additionally, with the introduction of set $\mathcal{S}[k]$, we define a new set of ``Long-Term Remaining'' nodes as $\mathcal{R}'[k]$, and discard the original ``Remaining'' set $\mathcal{R}[k]$. 

\textbf{Departing Soon.} 
This subset is denoted by $\mathcal{S}[k]$.
If $\rho_j^l[k] \leq \bar{\tau}_{j}$, then it comprises every node $v_j$ that is active at time step $k$ and will leave the network within the time interval $\{ k + \rho_j^l[k], ..., k+ \rho_j^u[k] \}$, where $\rho_j^u[k] > \rho_j^l[k] \geq 1$. 
It is defined as 
\begin{equation}\label{departing_soon_set_defn}
    \begin{aligned}
    \mathcal{S}[k] = & \{ v_j \in \mathcal{V}[k] \ | \  \exists\, s \in \{ k + \rho_j^l[k], ..., k+ \rho_j^u[k] \},\\
    & \ \text{ such that } v_j \in \mathcal{D}[s] \} . 
\end{aligned}
\end{equation}
We now define the set of ``Long-Term Remaining'' nodes. 

\textbf{Long-Term Remaining.}
This subset is denoted by $\mathcal{R}'[k]$.
If $\rho_j^l[k] > \bar{\tau}_{j}$, then it comprises every node $v_j$ that is active at time step $k$, remains active at time step $k+1$, and is not scheduled to depart within the interval $\{k+2, \ldots, k + \rho_j^l[k]-1 \}$. 
It is defined as 
\begin{equation}\label{remaining_updated_set_defn}
    \begin{aligned}
    \mathcal{R}'[k] = & \{ v_j \in \mathcal{V}[k] \cap \mathcal{V}[k+1], \ \  \text{and}\\
    & \ \ v_j \notin \mathcal{D}[s], \forall\, s \in \{k+2, \ldots, k + \rho_j^l[k]-1 \} \} . 
\end{aligned}
\end{equation}

Nodes operate according to the previously described modes: ``Long-Term Remaining,'' ``Arriving,'' ``Departing Soon,'' and ``Departing,'' transitioning between states as follows. 
At each time step $k$, each inactive node $v_j$ may transition to the ``Arriving'' mode to become active. 
Each active node $v_j$ scheduled to leave precisely at time step $k+1$ transitions to the ``Departing'' mode. 
Each active node $v_j$ that will leave at some time $s$ with $s \in \{ k + \rho_j^l[k], \ldots, k+ \rho_j^u[k] \}$ for which it holds $\rho_j^l[k] \leq \bar{\tau}_{j}$, transitions to the ``Departing Soon'' mode. 
Nodes that remain active across both time steps $k$ and $k+1$ and will not leave at some time $s$ with $s \in \{k+2, \ldots, k + \rho_j^l[k]-1 \}$ for which it holds $\rho_j^l[k] > \bar{\tau}_{j}$, operate in the ``Long-Term Remaining'' mode.

\AR{The above two subsets of nodes, together with the set of departing nodes $\mathcal{D}[k]$, split the set of active nodes $\mathcal{V}[k]$ into mutually exclusive and exhaustive subsets. 
Note that $\rho_j^l[k], \rho_j^u[k]$ are defined only for nodes $v_j \in \mathcal{V}[k] \setminus \mathcal{D}[k]$ (i.e., nodes that remain active through time step $k+1$); since $\rho_j^l[k] \geq 1$, the classification into $\mathcal{S}[k]$ or $\mathcal{R}'[k]$ concerns only future departures at $s \geq k+1$, and is therefore, by construction, disjoint from $\mathcal{D}[k]$.
Specifically, at each time step $k$, every active node $v_j \in \mathcal{V}[k]$ falls into exactly one of three cases: 
\begin{enumerate}
    \item $v_j \in \mathcal{D}[k]$: node $v_j$ departs from the network;
    \item $v_j \in \mathcal{S}[k]$: node $v_j$ is scheduled to possibly depart within the maximum delay horizon ($\rho_j^l[k] \le \bar{\tau}_j$);
    \item $v_j \in \mathcal{R}'[k]$: node $v_j$ remains active beyond the maximum delay horizon ($\rho_j^l[k] > \bar{\tau}_j$).
\end{enumerate}
Since $\rho_j^l[k] \le \bar{\tau}_j$ and $\rho_j^l[k] > \bar{\tau}_j$ are mutually exclusive and collectively exhaustive, we have
\[
\mathcal{V}[k] = \mathcal{D}[k] \cup \mathcal{S}[k] \cup \mathcal{R}'[k], 
\]
with 
\[\mathcal{D}[k] \cap \mathcal{S}[k] = \emptyset, \quad \mathcal{D}[k] \cap \mathcal{R}'[k] = \emptyset, \quad \mathcal{S}[k] \cap \mathcal{R}'[k] = \emptyset.
\] 
Thus, the operating modes form a complete partition of the active node set $\mathcal{V}[k]$ at every time step $k$, ensuring that every active node belongs to exactly one operating mode.}

Following the definition of the ``Departing Soon'' and ``Long-Term Remaining'' node subsets, we establish the following additional assumption that is necessary for our algorithm's operation. 
This assumption, together with Assumptions~\ref{assum_feedback_channel}, \ref{existence_stable_time_step}, \ref{strong_connectivity_stable_union_graph}, \ref{assum:bounded_message_delay}, and \ref{assum:know_when_it_leave}, forms the complete set of requirements for our proposed algorithm.

\begin{assumption}\label{awareness_of_long_termremaining_out_neighbors}
At every time step $k \geq 0$, each active node $v_j \in \mathcal{V}[k]$ knows the set of its long-term remaining out-neighbors $\mathcal{N}^{+}_j[k] \cap \mathcal{R}'[k]$. 
\end{assumption} 

Assumption~\ref{awareness_of_long_termremaining_out_neighbors} ensures that each active node $v_j \in \mathcal{V}[k]$ maintains knowledge of its long-term remaining out-neighbors $\mathcal{N}^{+}_j[k] \cap \mathcal{R}'[k]$ at every time step $k \geq 0$. 
This assumption is crucial for ensuring that each node transmits only to nodes that are currently active and will remain active for a duration exceeding their own maximum processing delay (i.e., if $v_j \in \mathcal{R}'[k]$, then $v_j$ will remain active for at least $\bar{\tau}_j$ time steps). 
By maintaining activity for a period exceeding its maximum processing delay, each receiving node can complete the processing of received information and forward it to another long-term remaining node before departing the network (thereby preventing information loss). 
Additionally, this assumption guarantees that departing nodes (i.e., nodes becoming inactive at the next time step) can successfully transmit their stored information to long-term remaining out-neighbor nodes ensuring no information is lost upon departure. 
Within our directed communication network, this assumption is realized through mechanisms similar to those employed in Assumption~\ref{awareness_of_remaining_out_neighbors}. 
Specifically, nodes utilize acknowledgment messages transmitted over the dedicated error-free narrowband feedback channel (see Assumption~\ref{assum_feedback_channel}) to communicate their operational status. 
This enables transmitting nodes to select only out-neighbors belonging to the ``Long-Term Remaining'' subset.

\subsection{Distributed Quantized Averaging for OMAS with Dynamic Communication Links and Processing Delays}\label{subsec_distr_alg_open_dynamic_procdelays}

Let $\mathcal{G}_d[k] = (\mathcal{V}[k], \mathcal{E}[k])$ denote an OMAS with dynamic communication links. 
Each node $v_j \in \mathcal{V}'$ is initialized with a quantized state $x_j \in \mathbb{Z}$. 
Additionally, Assumptions~\ref{assum_feedback_channel}, \ref{existence_stable_time_step}, \ref{strong_connectivity_stable_union_graph}, \ref{assum:bounded_message_delay}, \ref{assum:know_when_it_leave}, \ref{awareness_of_long_termremaining_out_neighbors} hold. 
At each time step $k$, each node maintains state, mass, and transmission variables (as denoted in the description of QAOD in Section~\ref{subsec_distr_alg_open_dynamic}). 
During the execution of our algorithm, each node executes the following operations (note that to avoid redundancy, we only mention the differences compared to the operations in QAOD).

\textbf{Long-Term Remaining Strategy.} 
During every time step $k$, each active node with long-term remaining status $v_j \in \mathcal{R}'[k]$ assigns to each of its outgoing edges $m_{lj}$ (including a virtual self-edge) a nonzero probability 
\begin{equation}\label{eq:longterm_remain_trans_prob}
     b_{lj}[k] \hspace{-.1cm} = \hspace{-.1cm} \left\{ 
\begin{aligned} 
\frac{1}{1 + |\mathcal{N}^{+}_j[k] \cap \mathcal{R}'[k]|} , \ \hspace{-.05cm} & v_l \in (\mathcal{N}^{+}_j[k] \cap \mathcal{R}'[k]) \cup \{v_j\}, \\
0, & \ \text{otherwise}.  
\end{aligned}
\right.
\end{equation}
Then, the rest of the operation is the same as the ``Remaining Strategy'' of QAOD. 

\textbf{Arriving Strategy.} 
Same as the ``Arriving Strategy'' of QAOD. 

\textbf{Departing Strategy.} 
When node $v_j$ departs from the network at time step $k$ (i.e., $v_j \in \mathcal{D}[k]$), it assigns a nonzero probability value $b_{lj}[k]$ to each of its outgoing edges $m_{lj}$ where $v_l \in \mathcal{N}^{+}_j[k] \cap \mathcal{R}'[k]$ according to: 
\begin{equation}\label{eq:depart_trans_prob_processdelay}
     b_{lj}[k]=\left\{ 
\begin{aligned} 
\frac{1}{|\mathcal{N}^{+}_j[k] \cap \mathcal{R}'[k]|}, & \quad v_l \in \mathcal{N}^{+}_j[k] \cap \mathcal{R}'[k], \\
0, & \quad v_l \notin \mathcal{N}^{+}_j[k] \cap \mathcal{R}'[k]. 
\end{aligned}
\right.
\end{equation}
Subsequently, it randomly selects a long-term remaining out-neighbor $v_l \in \mathcal{N}^{+}_j[k] \cap \mathcal{R}'[k]$ according to $b_{lj}[k]$ and transmits to $v_l$ its negative twice initial state $-2x_j$ and $-2r_j$ combined with its mass variables $y_j[k]$ and $z_j[k]$. 
The computation of transmission variables $c_{lj}^y[k]$ and $c_{lj}^z[k]$ follows the same procedure as described in~\eqref{trans_vari}.

\textbf{Departing Soon Strategy.}
Each active node $v_j$ with departing soon status at time step $k$ (i.e., $v_j \in \mathcal{S}[k]$), notifies its in-neighbors of its status to prevent them from transmitting information to it that may not be processed in time. 
During this phase, node $v_j$ focuses on processing any stored information it has received from previous time steps and prepares for its eventual departure. 
Nodes in the departing soon mode process the already received information (if any was received before its status change) but do not perform any transmissions to other nodes, as they will perform transmissions during the departure phase. 
In this way it is ensured that when the node eventually transitions to the departing mode at a future time step, it will be ready to transmit its processed information in order to remove its contribution from the network while preserving all useful information for the remaining nodes. 

The complete procedure for a single-step is detailed in QAPOD below. 
It describes the operational strategies for each node $v_j$ in all possible modes.

\begin{varalgorithm}{2} 
\caption{QAPOD}
\noindent \textbf{Input.} 
An OMAS with dynamic communication links $\mathcal{G}_d[k]=(\mathcal{V}[k], \mathcal{E}[k])$ with $n = | \mathcal{V}^\prime |$ nodes potentially participating, $n[k] = | \mathcal{V}[k] |$ active nodes, and $m[k] = | \mathcal{E}[k] |$ edges at each time step $k$. 
Each potentially participating node $v_j \in \mathcal{V}^\prime$ has initial state $x_j \in \mathbb{Z}$, and $r_j = 1$. 
Also, Assumptions~\ref{assum_feedback_channel}, \ref{existence_stable_time_step}, \ref{strong_connectivity_stable_union_graph}, \ref{assum:bounded_message_delay}, \ref{assum:know_when_it_leave}, \ref{awareness_of_long_termremaining_out_neighbors} hold.  
\\
\textbf{Initialization.} 
Same as QAOD. \\ 
\textbf{Iteration.} For each time step $k= 0, 1, 2, \dots$\\
\textbf{Arriving:} Same as QAOD. 
\\
\textbf{Departing Soon:} Each node $v_j \in \mathcal{S}[k]$ notifies
its in-neighbors of its operating mode. 
Also, it processes any stored information it has received (from previous time steps) preparing for its departure, and performs no transmissions while it operates in this mode. 
\\
\textbf{Departing:} Each node $v_j \in \mathcal{D}[k]$ does:
\begin{list4} 
\item[$1)$] Assigns a nonzero probability $b_{lj}[k]$ to each of its outgoing edges $m_{lj}$ 
as in \eqref{eq:depart_trans_prob_processdelay}. 
\item[$2)$] Chooses randomly one remaining out-neighbor $v_l \in \mathcal{N}^{+}_j[k] \cap \mathcal{R}'[k]$ according to probability $b_{lj}[k]$. 
\item[$3)$] Same as QAOD. 
\item[$4)$] Same as QAOD. 
\end{list4} 
\textbf{Long-Term Remaining:} Each node $v_j \in \mathcal{R}'[k]$ does:
\begin{list4} 
\item[$1)$] Assigns to each of its outgoing edges $m_{lj}$ (including a virtual self-edge) a nonzero probability $b_{lj}[k]$ as in $\eqref{eq:longterm_remain_trans_prob}$. 
\item[$2)$] Same as QAOD.  
\item[$3)$] Sets $c_{lj}^y[k] = 0$, $c_{lj}^z[k] = 0$, for every $v_l \in (\mathcal{N}^{+}_j[k] \cap \mathcal{R}'[k]) \cup \{v_j\}$. 
\item[$4)$] Same as QAOD.  
\item[$5)$] Same as QAOD.  
\item[$6)$] Same as QAOD.  
\begin{list4a}
\item[$6.1)$] Same as QAOD.  
\item[$6.2)$] Chooses $v_l \in (\mathcal{N}^{+}_j[k] \cap \mathcal{R}'[k]) \cup \{v_j\}$ according to $b_{lj}[k]$.  
\item[$6.3)$] Same as QAOD.   
\item[$6.4)$] Same as QAOD.  
\end{list4a}
\item[$7)$] Transmits $c_{lj}^y[k]$ and $c_{lj}^z[k]$, to $v_l$ for every $v_l \in (\mathcal{N}^{+}_j[k] \cap \mathcal{R}'[k]) \cup \{v_j\}$. 
\item[$8)$] Same as QAOD.  
\end{list4}
\textbf{Output:} Every active node $v_j \in \mathcal{V}[k]$ addresses \textbf{P1}. 
\label{algorithm2}
\end{varalgorithm}

\subsubsection{Formal Delayed-Message Buffer Model}

\AR{Regarding a formal delayed-message queue model, let us introduce the pending (unprocessed) buffered mass $\hat{y}_j[k], \hat{z}_j[k]$, denoting the portion of received mass that node $v_j$ has not yet incorporated into its state variables $y_j[k], z_j[k]$ at time step $k$. 
Let $y_j^{p}[k] \leq \hat{y}_j[k]$ and $z_j^{p}[k] \leq \hat{z}_j[k]$ denote the portions of the buffered mass processed and incorporated at time step $k$. The buffer then evolves as
\[
\hat{y}_j[k+1] = \hat{y}_j[k] - y_j^{p}[k] + y_j^{\mathrm{recv}}[k+1], 
\] 
\[ 
\hat{z}_j[k+1] = \hat{z}_j[k] - z_j^{p}[k] + z_j^{\mathrm{recv}}[k+1],
\]
where $y_j^{\mathrm{recv}}[k+1], z_j^{\mathrm{recv}}[k+1]$ denote the mass newly received at time step $k+1$, and the state variables are updated as $y_j[k+1] = y_j[k] + y_j^{p}[k]$, $z_j[k+1] = z_j[k] + z_j^{p}[k]$. 
Crucially, the processing schedule satisfies the following deadline constraint: any mass received at time step $k'$ must be fully processed (i.e., incorporated via $y_j^{p}[k''], z_j^{p}[k'']$ for some $k'' \in \{k', \dots, k' + \bar{\tau}_j\}$) no later than $\bar{\tau}_j$ time steps after its arrival. 
This constraint formalizes the delay bound $\tau_j[k] \leq \bar\tau_j$ from Assumption~\ref{assum:bounded_message_delay}, and ensures the buffer $\hat{y}_j[k], \hat{z}_j[k]$ is provably drainable within a bounded horizon. 
Upon transitioning to ``Departing Soon'' mode, node $v_j$ stops accepting new messages (i.e., $y_j^{\mathrm{recv}}[k+1] = z_j^{\mathrm{recv}}[k+1] = 0$ for all subsequent time steps until departure) and fully empties its buffer (i.e., $y_j^p[k] = \hat y_j[k]$, $z_j^p[k] = \hat z_j[k]$ within at most $\bar\tau_j$ steps), before transmitting its processed mass to a Long-Term Remaining out-neighbor at departure. This guarantees $\hat{y}_j[k] = \hat{z}_j[k] = 0$ at the time of departure, which is the key property ensuring exact mass preservation.}

\subsection{Correctness Analysis of QAPOD}

We now establish the correctness of our QAPOD via the following theorem. 

\begin{theorem}\label{main_convergence_condition_theorem_process_delays}
Let us consider an OMAS with dynamic communication links $\mathcal{G}_d[k]=(\mathcal{V}[k], \mathcal{E}[k])$ with $n = | \mathcal{V}^\prime |$ nodes potentially participating, $n[k] = | \mathcal{V}[k] |$ active nodes, and $m[k] = | \mathcal{E}[k] |$ edges at each time step $k$.  
Assumptions~\ref{assum_feedback_channel}, \ref{existence_stable_time_step}, \ref{strong_connectivity_stable_union_graph}, \ref{assum:bounded_message_delay}, \ref{assum:know_when_it_leave}, \ref{awareness_of_long_termremaining_out_neighbors} hold 
Let us also assume that all potentially participating nodes execute QAPOD. 
There exists a time step $k_0$ such that for every active node $v_j \in \mathcal{V}[k]$ we have 
$$ 
( q^s_j[k] = \lfloor q[k] \rfloor ) \ \text{or} \ ( q^s_j[k] = \lceil q[k] \rceil ) , 
$$ 
for $k \geq k_0$, where $q[k]$ is defined in \eqref{goal}, if and only if for every departing node $v_j \in \mathcal{D}[k]$ it holds that 
\begin{equation}\label{condition_for_correctness_process_delays} 
    | \ \mathcal{N}^{+}_j[k] \cap \mathcal{R}'[k] \ | \geq 1 \ , 
\end{equation} 
for every time step $k$. 
\end{theorem}

\textbf{PROOF.} See Appendix~\ref{convergence_Alg2_prob_P1_process}. \hspace*{\fill} $\square$

% \begin{proof} 
% See Appendix~\ref{convergence_Alg2_prob_P1_process}.
% \end{proof}

\begin{remark}[Intuition of Theorem~\ref{main_convergence_condition_theorem_process_delays}]
In Theorem~\ref{main_convergence_condition_theorem_process_delays} we showed that the correctness of QAPOD under processing delays relies on every departing node having at least one long-term remaining out-neighbor. 
This ensures information is not lost when departures occur. 
The proof structure is similar to Theorem~\ref{main_convergence_condition_theorem}, but with an important modification. 
Nodes classified as ``Remaining'' during the operation of QAOD, are now split into two categories: ``Long-Term Remaining'' and ``Departing Soon.''
Following this framework, the operation of QAPOD closely mirrors that of QAOD. 
However, the crucial difference is that nodes in the departing soon mode remain active in the network but do not transmit or receive new messages. 
Since these departing soon nodes have not yet exited the network, all stored information remains intact, and no loss occurs. 
Therefore, throughout the execution of QAPOD, information is preserved. 
Overall, departing soon nodes act as temporary holders of information until they transition into the departing mode, at which point they transmit to a long-term remaining node their stored information to ensure that the sum preservation property still holds. 
\end{remark}

% % ===============================================
% %
% %
% % ALGORITHM 3 
% %
% %
% % ===============================================

\section{Quantized Averaging over Indefinitely OMAS with Dynamic Communication Links}\label{sec:always_open_dyn_net}

% In many practical applications, networks may never reach a stable configuration \cite{deplano2025optimization}. 
% Let us consider for example distributed machine learning systems for mobile applications. 
% Devices may join and leave the network based on user behavior, or connectivity availability, creating an indefinitely dynamic environment. 
% In such scenarios, Assumption~\ref{existence_stable_time_step} may not reflect real-world conditions. 
% Additionally, when nodes depart unexpectedly (due to failures or resource limitations) their local data remain valuable for the global objective, and losing this information could significantly impact the quality of the final result. 
% Overall, including the local data of all nodes that have ever participated in the network enables the remaining nodes to compute more accurate and reliable aggregate results. 
% To address these challenges, it becomes essential to design algorithms that can operate effectively in networks that remain open indefinitely, while preserving the contributions of all nodes that have ever participated in the computation, regardless of their current activity status.

In many practical applications, networks may never reach a stable configuration \cite{deplano2025optimization}. 
For example, in distributed machine learning systems for mobile applications, devices continuously join and leave based on user behavior and connectivity. 
This creates indefinitely dynamic environments where Assumption~\ref{existence_stable_time_step} does not hold. 
Additionally, when nodes depart unexpectedly (e.g., due to resource limitations) their local data may remain valuable for the global objective, and losing this information could substantially degrade the quality of the final solution.
Therefore, utilizing the local data of all historically active nodes enables the remaining nodes to compute more accurate and reliable results. 
Motivated by the aforementioned challenges, we propose a distributed algorithm (detailed below as QAIOD) that addresses problem \textbf{P2}. 
For our proposed algorithm, Assumptions~\ref{assum_feedback_channel}, \ref{awareness_of_remaining_out_neighbors}, and \ref{no_delays_assum} are required. 
Assumptions~\ref{existence_stable_time_step}, \ref{strong_connectivity_stable_union_graph}, \ref{assum:bounded_message_delay}, and \ref{assum:know_when_it_leave} do not apply. 
Specifically, Assumptions~\ref{existence_stable_time_step}, \ref{strong_connectivity_stable_union_graph} do not apply because we have an indefinitely open multi-agent system (i.e., nodes may depart and/or arrive in the network during \textit{every} time step $k$). 
Furthermore, Assumptions~\ref{assum:bounded_message_delay}, and \ref{assum:know_when_it_leave} do not apply since we do not assume processing delays in this section (note that extending QAIOD to operate in the presence of processing delays is feasible adopting the strategies we presented in Section~\ref{sec:open_dyn_network_process_del}, but for simplicity of analysis we assume that the processing performed by each node experiences no delay). 
Additionally, we also impose the following assumption that is essential for our subsequent analysis. 

\begin{assumption}[Open $T'$-jointly strong connectivity]\label{open_strong_connectivity_stable_union_graph}
Let us consider an OMAS with dynamic communication links $\mathcal{G}_d[k] = (\mathcal{V}[k], \mathcal{E}[k])$. 
During any time step $k$, for $L \in \mathbb{N}$, let us define the subset of nodes 
\begin{equation}\label{subset_in_L_nodes}
    \mathcal{I}^L[k] = \bigcup_{t=k}^{k+L} \mathcal{V}[t] , 
\end{equation} 
denoting all nodes that were active for at least one time step in the network during the time interval $\{ k, ..., k + L \}$. 
Additionally, let us define the subset of edges 
\begin{equation}\label{subset_in_L_edges}
    \mathcal{Q}^L[k] = \{ (v_j, v_i) \in \mathcal{E} \ | \ v_j, v_i \in \mathcal{I}^L[k] \} 
\end{equation}
denoting all possible edges in the network among all nodes that were active for at least one time step during the time interval $\{ k, ..., k + L \}$ (recall from Section~\ref{subsec_open_dynam_networks} that $\mathcal{E}$ denotes the set of all possible edges in the network). 
Let us suppose now that for every time step $k$ there exists $L' \in \mathbb{N}$ such that the \textit{anytime virtual union digraph} defined as $\widehat{\mathcal{G}}_d[k] = (\mathcal{I}^{L'}[k], \mathcal{Q}^{L'}[k])$ is strongly connected. 
For every time step $k$ we have that there exists $T' \in \mathbb{N}$, where $T' \geq L'$, such that for the \textit{anytime union digraph} defined as $\widehat{\mathcal{G}}^{ \{ 1, ..., T' \}}_d[k] = (\mathcal{I}^{T'}[k], \cup_{t= k}^{k + T'} \mathcal{E}[t])$ it holds 
\begin{equation}\label{open_strong_conn_equations}
\begin{aligned} 
    \mathcal{I}^{T'}[k] = & \ \mathcal{I}^{L'}[k], \ \  \text{and} \\
    \cup_{t= k}^{k + T'} \mathcal{E}[t] = & \ \mathcal{Q}^{L'}[k] \ . 
\end{aligned}
\end{equation}
This property means that the OMAS with dynamic communication links $\mathcal{G}_d[k]$ is \textit{Open} $T'$-\textit{jointly strongly connected} for any time step $k$.  
\end{assumption}

% Assumption~\ref{open_strong_connectivity_stable_union_graph} implies that for every time step $k$, there exists a finite interval $\{ k, ..., k+T' \}$ (where $T' \in \mathbb{N}$) over which all possible edges among nodes that were active at least once during this interval appear at least once, and they form a strongly connected network. 
% This assumption ensures that there exists at least one directed path between every pair of active nodes infinitely often. 
% As a result, information originating from any active node is guaranteed to eventually reach every other active node.
Assumption~\ref{open_strong_connectivity_stable_union_graph} implies that for every time step $k$, there exists a finite interval $\{ k, ..., k+T' \}$ (where $T' \in \mathbb{N}$) over which all possible edges among nodes that were active at least once during this interval appear at least once, and they form a strongly connected network. 
\AR{It does not require instantaneous strong connectivity among all potentially participating nodes at any single time step. 
It only requires that, over each recurring interval, the edges among nodes that have been active at least once during that interval collectively form a strongly connected virtual union digraph. 
This accommodates indefinitely open networks such as peer-to-peer systems or mobile ad hoc networks, where nodes intermittently join, leave, and reactivate, and no fixed subset of nodes need ever be simultaneously connected. 
This condition is naturally satisfied without centralized coordination, e.g., whenever active agents intermittently encounter each other or relay packets through underlying multi-hop routing protocols over bounded recurring windows. 
In our analysis below, we show that since the set of potentially participating nodes $\mathcal{V}'$ is finite, the historical active set $\mathcal{H}[k]$ stabilizes after a finite time step $\iota''$. After time step $\iota''$, the recurrent connectivity in Assumption~\ref{open_strong_connectivity_stable_union_graph} ensures that the preserved historical mass continues to propagate and mix across the network at every subsequent time step, driving the quantized state of every active node toward the fixed historical average and thereby guaranteeing finite-time convergence.}

\subsection{Distributed Quantized Averaging Algorithm over Indefinitely OMAS with Dynamic Communication Links}\label{subsec_distr_alg_open_dynamic_always}

Let $\mathcal{G}_d[k] = (\mathcal{V}[k], \mathcal{E}[k])$ denote an OMAS with dynamic communication links. 
Each node $v_j \in \mathcal{V}'$ is initialized with a quantized state $x_j \in \mathbb{Z}$. 
Additionally, each node $v_j$ initializes the participation variable $\eta_j = 1$ (this variable is set to $1$, until $v_j$ becomes active at least once, and updated to $0$ once this occurs), and its state update variable $\nu_j = -1$ (this variable stores the last time step at which node $v_j$ updated its state). 
At every time step $k$, all nodes maintain state, mass, and transmission variables (as described in QAOD in Section~\ref{subsec_distr_alg_open_dynamic}).
During the execution of our algorithm, each node executes the following operations (we only mention the differences compared to the operations in QAOD, omitting the common steps for brevity).

\textbf{Remaining Strategy.} 
During every time step $k$, each active node with remaining status $v_j \in \mathcal{R}[k]$ sets $\eta_j = 0$. 
Then, it assigns to each of its outgoing edges $m_{lj}$ (including a virtual self-edge) a nonzero probability following \eqref{eq:remain_trans_prob}. 
If $z_j[k] = 0$, it does not update its state variables, its state update variable $\nu_j$, and its transmission variables (i.e., it performs no transmissions to its out-neighbors). 
If $z_j[k] > 1$, sets $\nu_j = k$ and its operation is the same as the ``Remaining Strategy'' of QAOD.

% Then, if $z_j[k] \geq 1$, it updates its state variables to be equal to the mass variables and also updates its transmission variables, and sets $\nu_j = k$. 
% For updating its transmission variables $c_{lj}^y[k]$, $c_{lj}^z[k]$, it splits $y_j[k]$ into $z_j[k]$ equal pieces, keeps one piece for itself and transmits the other pieces to randomly chosen out-neighbors or itself according to the assigned nonzero probabilities $b_{lj}[k]$ above. 

% Then, $v_j$ receives the transmission variables from every in-neighbor $v_i \in \mathcal{N}^{-}_j[k]$, and updates its mass variables $y_j[k+1]$, $z_j[k+1]$ following \eqref{eq:mass_var_update}. 

\textbf{Departing Strategy.}  
When node $v_j$ departs from the network at time step $k$ (i.e., $v_j \in \mathcal{D}[k]$), it assigns a nonzero probability $b_{lj}[k]$ to each outgoing edge $m_{lj}$ with $v_l \in \mathcal{N}_j^+[k] \cap \mathcal{R}[k]$ following \eqref{eq:depart_trans_prob}.  
It then randomly selects a remaining out-neighbor $v_l \in \mathcal{N}_j^+[k] \cap \mathcal{R}[k]$ according to $b_{lj}[k]$ and transmits its stored mass variables $y_j[k]$ and $z_j[k]$ to $v_l$. 
The transmission variables are computed as:  
\begin{equation}\label{trans_vari_alwaysopen}
\begin{aligned}
    c_{lj}^y[k] &= y_j[k],\\
    c_{lj}^z[k] &= z_j[k].
\end{aligned}
\end{equation}

\textbf{Arriving Strategy.}
When node $v_j$ arrives in the network at time step $k$ (i.e., $v_j \in \mathcal{A}[k]$), it checks the value of its participation variable $\eta_j$.  
If $v_j$ was previously active (i.e., $\eta_j = 0$), it updates its state variables $y_j^s[k+1]$, $z_j^s[k+1]$, $q_j^s[k+1]$, and mass variables $y_j[k+1]$, $z_j[k+1]$ as follows:
\begin{equation}\label{init_variables_alwaysopen}
\begin{aligned}
    y_j[k+1] &= 0, & z_j[k+1] &= 0, \\
    y_j^s[k+1] &= y_j^s[\nu_j], & z_j^s[k+1] &= z_j^s[\nu_j], \\
    q_j^s[k+1] &= q_j^s[\nu_j]. & &
\end{aligned}
\end{equation}
Then it proceeds to interact with its active neighbors in the following time step. 
If $v_j$ becomes active for the first time (i.e., $\eta_j = 1$), then its operation is the same as the ``Arriving Strategy'' of QAOD.

\AR{Note here that at each time step $k$, the per-step computation, communication, and storage of QAIOD are the same as those of QAOD.}

In QAIOD below we provide a summary of a single step of our proposed procedure, outlining all operational modes for each node $v_j$. 

\begin{varalgorithm}{3} 
\caption{QAIOD} 
\noindent \textbf{Input.} 
An OMAS with dynamic communication links $\mathcal{G}_d[k]=(\mathcal{V}[k], \mathcal{E}[k])$ with $n = | \mathcal{V}^\prime |$ nodes potentially participating, $n[k] = | \mathcal{V}[k] |$ active nodes, and $m[k] = | \mathcal{E}[k] |$ edges at each time step $k$. 
Each potentially participating node $v_j \in \mathcal{V}^\prime$ has initial state $x_j \in \mathbb{Z}$, and $r_j = 1$. 
Assumptions~\ref{assum_feedback_channel}, \ref{awareness_of_remaining_out_neighbors}, \ref{no_delays_assum}, and \ref{open_strong_connectivity_stable_union_graph} hold. 
\\ 
\textbf{Initialization.} 
Each node $v_j \in \mathcal{V}^\prime$ sets $y_j[0] = 2x_j$, $z_j[0] = 2r_j$, and $\eta_j = 1$, $\nu_j = -1$. 
\\ 
\textbf{Iteration.} For each time step $k= 0, 1, 2, \dots$\\
\textbf{Arriving:} Each node $v_j \in \mathcal{A}[k]$ does: 
\begin{list4} 
\item[$1)$] \textbf{If} $\eta_j = 0$ \textbf{then} it updates its state $y_j^s[k+1]$, $z_j^s[k+1]$, $q_j^s[k+1]$, and mass variables $y_j[k+1]$, $z_j[k+1]$ as \eqref{init_variables_alwaysopen}. 
\item[$2)$] \textbf{If} $\eta_j = 1$ \textbf{then} it updates its state $y_j^s[k+1]$, $z_j^s[k+1]$, $q_j^s[k+1]$, and mass variables $y_j[k+1]$, $z_j[k+1]$ as \eqref{init_variables}. 
\end{list4} 
\textbf{Departing:} Each node $v_j \in \mathcal{D}[k]$ does: 
\begin{list4} 
\item[$1)$] Assigns a nonzero probability $b_{lj}[k]$ to each of its outgoing edges $m_{lj}$ as in \eqref{eq:depart_trans_prob}. 
\item[$2)$] Chooses randomly one remaining out-neighbor $v_l \in \mathcal{N}^{+}_j[k] \cap \mathcal{R}[k]$ according to probability $b_{lj}[k]$. 
\item[$3)$] Computes transmission variables $c_{lj}^y[k]$, $c_{lj}^z[k]$ as in \eqref{trans_vari_alwaysopen}. 
\item[$4)$] Transmits $c_{lj}^y[k]$, $c_{lj}^z[k]$ to the selected remaining out-neighbor $v_l$. 
\end{list4} 
\textbf{Remaining:} Each node $v_j \in \mathcal{R}[k]$ does:
\begin{list4} 
\item[$1)$] Sets $\eta_j = 0$. 
\item[$2)$] Same as Step~$1$ of QAOD. 
\item[$3)$] \textbf{If} $z_j[k] > 1$ \textbf{then}
\begin{list4a} 
\item[$3.1)$] Same as Step~$2$ of QAOD. 
\item[$3.2)$] Sets $\nu_j = k$. 
\item[$3.3)$] Same as Step~$3$ of QAOD. 
\item[$3.4)$] Same as Step~$4$ of QAOD. 
\item[$3.5)$] Same as Step~$5$ of QAOD. 
\item[$3.6)$] Same as Step~$6$ of QAOD. 
\begin{list4a} 
\item Same as Steps~$6.1 - 6.4$ QAOD. 
\end{list4a} 
\end{list4a} 
\item[$4)$] Same as Step~$7$ of QAOD. 
\item[$5)$] Same as Step~$8$ of QAOD. 
\end{list4}
\textbf{Output:} Every active node $v_j \in \mathcal{V}[k]$ addresses \textbf{P2}. 
\label{algorithm3}
\end{varalgorithm}

\subsection{Correctness Analysis of QAIOD}

We now establish the correctness of our QAIOD via the following theorem. 

% \begin{theorem}\label{main_convergence_condition_theorem_alwaysopen}
% Let us consider an OMAS with dynamic communication links $\mathcal{G}_d[k]=(\mathcal{V}[k], \mathcal{E}[k])$ with $n = | \mathcal{V}^\prime |$ nodes potentially participating, $n[k] = | \mathcal{V}[k] |$ active nodes, and $m[k] = | \mathcal{E}[k] |$ edges at each time step $k$.  
% Suppose that Assumptions~\ref{assum_feedback_channel}, \ref{awareness_of_remaining_out_neighbors}, \ref{no_delays_assum}, and \ref{open_strong_connectivity_stable_union_graph} hold. 
% Let us also assume that all potentially participating nodes execute QAIOD. 
% There exists a time step $k_0$ such that for every active node $v_j \in \mathcal{V}[k]$ we have 
% $$ 
% ( q^s_j[k] = \lfloor q'[k] \rfloor ) \ \text{or} \ ( q^s_j[k] = \lceil q'[k] \rceil ) , 
% $$ 
% for $k \geq k_0$ where $q'[k]$ is defined in \eqref{eq:extend_goal}, if and only if for every departing node $v_j \in \mathcal{D}[k]$ it holds that 
% \begin{equation}\label{condition_for_correctness_alwaysopen} 
%     | \ \mathcal{N}^{+}_j[k] \cap \mathcal{R}[k] \ | \geq 1 \ , 
% \end{equation} 
% for every time step $k$. 
% \end{theorem}

\AR{
\begin{theorem}\label{main_convergence_condition_theorem_alwaysopen}
Let us consider an OMAS with dynamic communication links $\mathcal{G}_d[k]=(\mathcal{V}[k], \mathcal{E}[k])$ with $n = | \mathcal{V}^\prime |$ nodes potentially participating, $n[k] = | \mathcal{V}[k] |$ active nodes, and $m[k] = | \mathcal{E}[k] |$ edges at each time step $k$.  
Suppose that Assumptions~\ref{assum_feedback_channel}, \ref{awareness_of_remaining_out_neighbors}, \ref{no_delays_assum}, and \ref{open_strong_connectivity_stable_union_graph} hold. 
Since $|\mathcal{V}^\prime| = n < \infty$, let $\iota^{\prime \prime}$ be the finite time step after which no new nodes enter the network for the first time. 
% so that $\mathcal{H}[k] = \mathcal{H}_{\mathrm{final}}$ and $q'[k] = q'_{\mathrm{final}} \triangleq \frac{1}{|\mathcal{H}_{\mathrm{final}}|} \sum_{v_j \in \mathcal{H}_{\mathrm{final}}} x_j$ for all $k \geq \iota^\prime$.
Let us also assume that all potentially participating nodes execute QAIOD. 
There exists a finite time step $k_0 \geq \iota^{\prime \prime}$ such that for every active node $v_j \in \mathcal{V}[k]$ we have 
$$ 
( q^s_j[k] = \lfloor q'[k] \rfloor ) \ \text{or} \ ( q^s_j[k] = \lceil q'[k] \rceil ) , 
$$ 
for all $k \geq k_0 \geq \iota^{\prime \prime}$, if and only if for every departing node $v_j \in \mathcal{D}[k]$ it holds that 
\begin{equation}\label{condition_for_correctness_alwaysopen} 
    | \ \mathcal{N}^{+}_j[k] \cap \mathcal{R}[k] \ | \geq 1 \ , 
\end{equation} 
for every time step $k$. 
\end{theorem}
}

\textbf{PROOF.} See Appendix~\ref{convergence_Alg3_prob_P2}. \hspace*{\fill} $\square$

% \begin{proof} 
% See Appendix~\ref{convergence_Alg3_prob_P2}.
% \end{proof}

\AR{\begin{remark}
Our Theorem~\ref{main_convergence_condition_theorem_alwaysopen} establishes that, each departing node must have at least one remaining out-neighbor for QAIOD to solve problem \textbf{P2}. 
However, since $\mathcal{V}'$ is finite, the historical active set $\mathcal{H}[k]$ is monotonically non-decreasing and bounded. 
Therefore, there exists a finite time step $\iota''$ after which no new nodes enter the network for the first time (i.e., $\mathcal{H}[k]=\mathcal{H}_{\mathrm{final}}$ for all $k\geq\iota''$).  
For $k<\iota''$, each first-time arrival shifts the target average $q'[k]$, and QAIOD tracks this by injecting the new node's initial state into the mass variables upon arrival (no fixed target exists yet, so no convergence claim is made). 
Once $k\geq\iota''$, $q'[k]$ settles to a fixed value, and Assumption~\ref{open_strong_connectivity_stable_union_graph} guarantees information from every historically active node propagates across recurring strongly connected topologies, driving $q^s_j[k]$ to fulfill \eqref{goal_P1_ext} in finite time. 
This is why our convergence guarantee in Theorem~\ref{main_convergence_condition_theorem_alwaysopen} is stated for $k\geq k_0\geq\iota''$. 
\end{remark}} 

\section{Application: Distributed Sensor Fusion for Environmental Monitoring}\label{results}

To validate the performance and practicality of our proposed algorithms, we consider a distributed sensor fusion scenario for environmental monitoring \cite{2019:Carminati_Marcuccio_44_52, 2018:Lombardo_Elsayed_1214_1222}.  
In this application, a network of sensors (e.g., deployed across a forest or urban area) is tasked with calculating the average value of an environmental quantity (e.g., temperature, air pollution, or radiation levels). 
The underlying network is open and dynamic. 
Specifically, sensors may join the network when activated (or deployed), leave due to battery depletion or damage, and communication links may vary due to environmental obstructions or mobility. 
Additionally, the limited bandwidth and energy resources of sensor nodes necessitate efficient quantized communication, and the nature of the environmental monitoring task demands finite-time convergence to enable timely decision-making. 

\subsection{Operational Comparisons of QAOD, QAPOD, QAIOD}\label{first_sec_results}

\textbf{Setup.}
We consider an OMAS with dynamic communication links $\mathcal{G}_d[k] = (\mathcal{V}[k], \mathcal{E}[k])$. 
Each sensor node $v_j \in \mathcal{V}'$ is initialized with a local quantized state $x_j \in \mathbb{Z}$ randomly chosen from the interval $[1, 10]$ with uniform probability. 
We focus on the three scenarios presented below. 
In each scenario we aim to calculate the current active set average (i.e., problem \textbf{P1}) executing QAOD and QAPOD, and also the average of all previously and currently active sensor nodes (i.e., problem \textbf{P2}) via QAIOD. 
For each scenario, we plot the time-varying error $\varepsilon[k]$, averaged over $100$ random digraphs of the corresponding size. 
The error $\varepsilon[k]$ is defined as 
\begin{equation} 
\begin{aligned} 
\varepsilon[k] = & \sum\limits_{\{v_j \in \mathcal{V}[k] | \lceil y_j[k] / z_j[k] \rceil > \lceil q_\text{alg}[k] \rceil\}} \!\!\!\!\!\!\!\!\!\!\!\!\!(\lceil 
 y_j[k] / z_j[k] \rceil \!-\! \lceil q_\text{alg}[k] \rceil) +\! \\&\sum\limits_{\{v_j \in \mathcal{V}[k] | \lfloor y_j[k] / z_j[k] \rfloor < \lfloor q_\text{alg}[k] \rfloor\}}\!\!\!\!\!\!\!\!\!\!\!\!\!(\lfloor q_\text{alg}[k] \rfloor - \lfloor y_j[k] / z_j[k] \rfloor) , 
\end{aligned}\label{error_plot}
\end{equation}
where for QAOD and QAPOD (problem \textbf{P1}) we set $q_\text{alg}[k] = q[k]$ (see \eqref{goal}), and for QAIOD (problem \textbf{P2}), we set $q_\text{alg}[k] = q'[k]$ (see \eqref{eq:extend_goal}). 
\\ \noindent 
\textit{Scenario~1.} 
In this scenario we demonstrate the correctness of our algorithms. 
We consider that $\mathcal{G}_d[k]$ has $n=150$ potentially participating nodes with $n[0]=100$ active nodes at time step $k=0$. 
At each time step $k$, the node departure and arrival rates are both set to $10\%$ of the current number of active nodes. 
The number of arriving nodes is then perturbed. 
With a $55\%$ probability, the number of arriving nodes is increased by $1$, and with a $45\%$ probability, it is decreased by $1$. 
Furthermore, if the number of inactive nodes is smaller than the $10\%$ of the current number of active nodes, all remaining inactive nodes join the network at the next time step. 
For QAOD and QAPOD we have: (i) during the time interval $1 < k < 80$ nodes may enter or leave the network, and $k^\prime = 80$ (see Assumption~\ref{existence_stable_time_step}), (ii) the network $\mathcal{G}_d[k]$ is $T$-jointly strongly connected with $T=20$ and (see Assumption~\ref{strong_connectivity_stable_union_graph}), and (iii) for every departing node $v_j \in \mathcal{D}[k]$ it holds that $| \ \mathcal{N}^{+}_j[k] \cap \mathcal{R}[k] \ | \geq 1$ and $| \ \mathcal{N}^{+}_j[k] \cap \mathcal{R}^\prime[k] \ | \geq 1$, respectively, during every time step $k$ (see Theorem~\ref{main_convergence_condition_theorem} and Theorem~\ref{main_convergence_condition_theorem_process_delays}). 
Additionally, the operation of QAPOD is demonstrated for maximum processing delay $\bar{\tau}=5$, and also $\bar{\tau}=10$.
For QAIOD we have (i) the network $\mathcal{G}_d[k]$ is open $T^\prime$-jointly strongly connected with $T^\prime=20$ (see Assumption~\ref{open_strong_connectivity_stable_union_graph}), and (ii) for every departing node $v_j \in \mathcal{D}[k]$ it holds that $| \ \mathcal{N}^{+}_j[k] \cap \mathcal{R}[k] \ | \geq 1$ for every time step $k$ (see Theorem~\ref{main_convergence_condition_theorem_alwaysopen}). 
\\ \noindent 
\textit{Scenario~2.} 
In this scenario we show how convergence scales with the network size. 
We consider three cases of initial number of active nodes $n[0]$ and number of potentially participating nodes $n$ of $\mathcal{G}_d[k]$: (i) $n[0]=100$ with $n=150$, (ii) $n[0]=250$ with $n=300$, and (iii) $n[0]=500$ with $n=600$. 
For all cases, QAPOD uses a maximum processing delay of $\bar{\tau} = 5$. 
All other parameters for QAOD, QAPOD, and QAIOD remain consistent with \textit{Scenario~1}. 
\\ \noindent 
\textit{Scenario~3.} 
This scenario evaluates the scaling of convergence with the node departure and arrival rate in a network of $n=150$ potentially participating nodes, initially with $n[0]=100$ active nodes. 
We consider two departure and arrival rates: (i) $10\%$ and (ii) $50\%$, of the current number of active nodes. 
Similarly to \textit{Scenario~1} the number of arriving nodes is perturbed. 
With a $55\%$ probability, it is increased by $1$, and with a $45\%$ probability, it is decreased by $1$. 
Furthermore, if the number of inactive nodes is smaller than the $10\%$ (or $50\%$ considering the corresponding departure rate) of the current number of active nodes, all remaining inactive nodes join the network at the next time step. 
In QAOD and QAPOD, the aforementioned departure rates govern node departures during time steps $1 < k < 80$ (with $k'=80$ from Assumption~\ref{existence_stable_time_step}). 
For QAIOD, they apply throughout its operation in the indefinitely open multi-agent system. 
QAPOD is configured with $\bar{\tau}=5$. 
All remaining parameters are consistent with \textit{Scenario~1}. 

\begin{figure}[t] 
\begin{center}
\includegraphics[width=8.4cm]{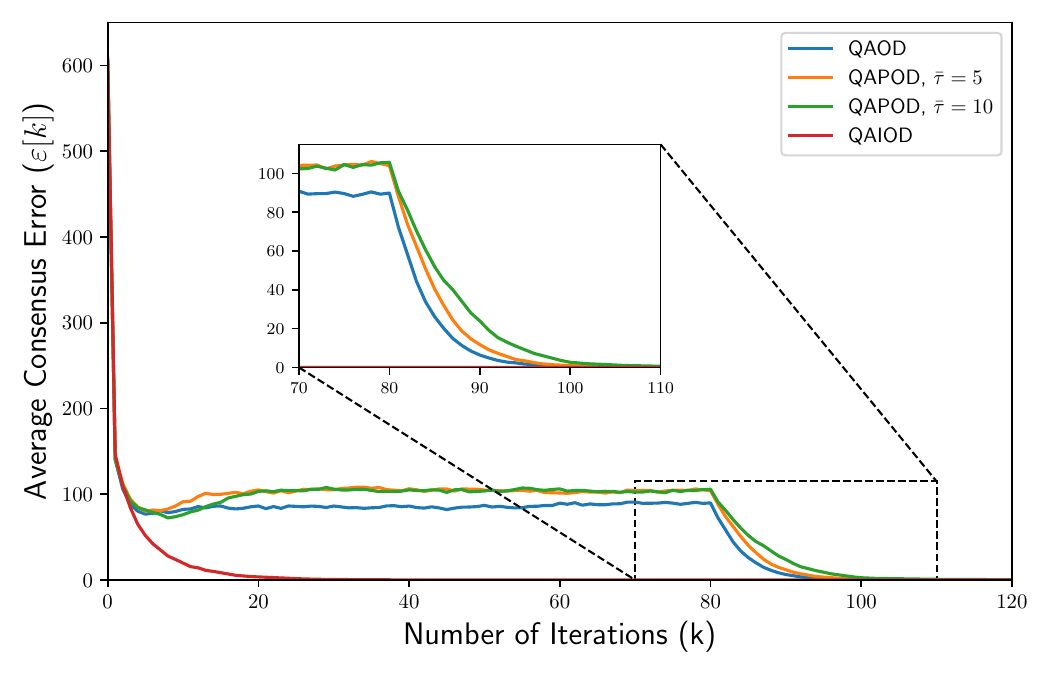} 
\caption{Evolution of error $\varepsilon[k]$ (shown in \eqref{error_plot}) over time $k$ averaged over $100$ random digraphs of $150$ potentially participating nodes, during the execution of QAOD, QAPOD, and QAIOD.}
\label{fig:scenario1} 
\end{center} 
\end{figure} 

In Fig.~\ref{fig:scenario1}, we observe that all our proposed algorithms achieve finite-time convergence. 
For QAOD, the node states track the quantized average but do not stabilize for $k < 80$ due to ongoing node arrivals and departures. 
However, for $k \geq 80$, the states converge to the exact quantized average of their initial states (driving the error $\varepsilon[k]$ to zero) after approximately $95$ time steps. 
The convergence characteristics of QAPOD for $k < 80$ are similar to those of QAOD. 
Nonetheless, QAPOD converges after approximately  $105$ and $110$ time steps for $\bar{\tau}=5$ and $\bar{\tau}=10$, respectively, demonstrating that a larger maximum processing delay increases the time to convergence.
For QAIOD, the active nodes converge to the average of all previously and currently active nodes (a result guaranteed by the network connectivity conditions in Assumption~\ref{open_strong_connectivity_stable_union_graph}). 
The results show that QAIOD achieves fast convergence after approximately $40$ time steps, demonstrating its robustness despite the dynamic nature of the indefinitely OMAS. 

\begin{figure}[t] 
\begin{center} 
\includegraphics[width=8.4cm]{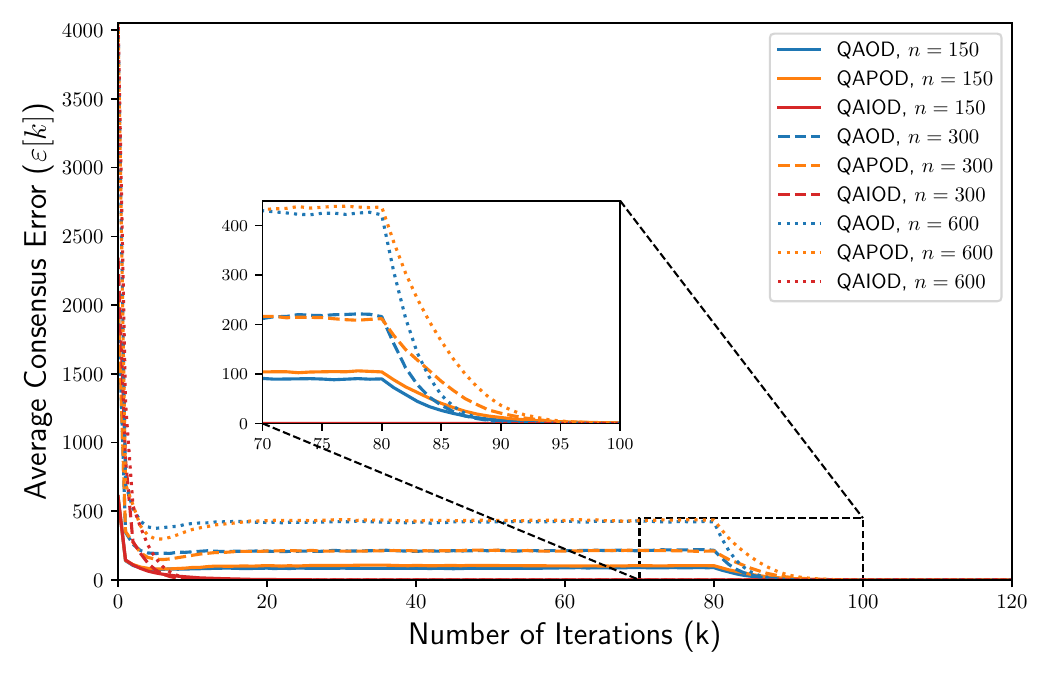} 
\caption{Evolution of error $\varepsilon[k]$ (shown in \eqref{error_plot}) over time $k$ averaged over $100$ random digraphs of $150, 300$, and $600$ potentially participating nodes, respectively, during the execution of QAOD, QAPOD, and QAIOD.} 
\label{fig:scenario2}
\end{center} 
\end{figure} 

In Fig.~\ref{fig:scenario2}, observations are consistent with those in Fig.~\ref{fig:scenario1}.  
While the behavior of $\varepsilon[k]$ varies with network size, our proposed algorithms demonstrate fast finite-time convergence for all network sizes.
In particular, QAOD and QAPOD converge after approximately $95-105$ time steps. 
QAIOD achieves convergence after approximately $40$ time steps, also independent of the network size. 

\begin{figure}[t] 
\begin{center} 
\includegraphics[width=8.4cm]{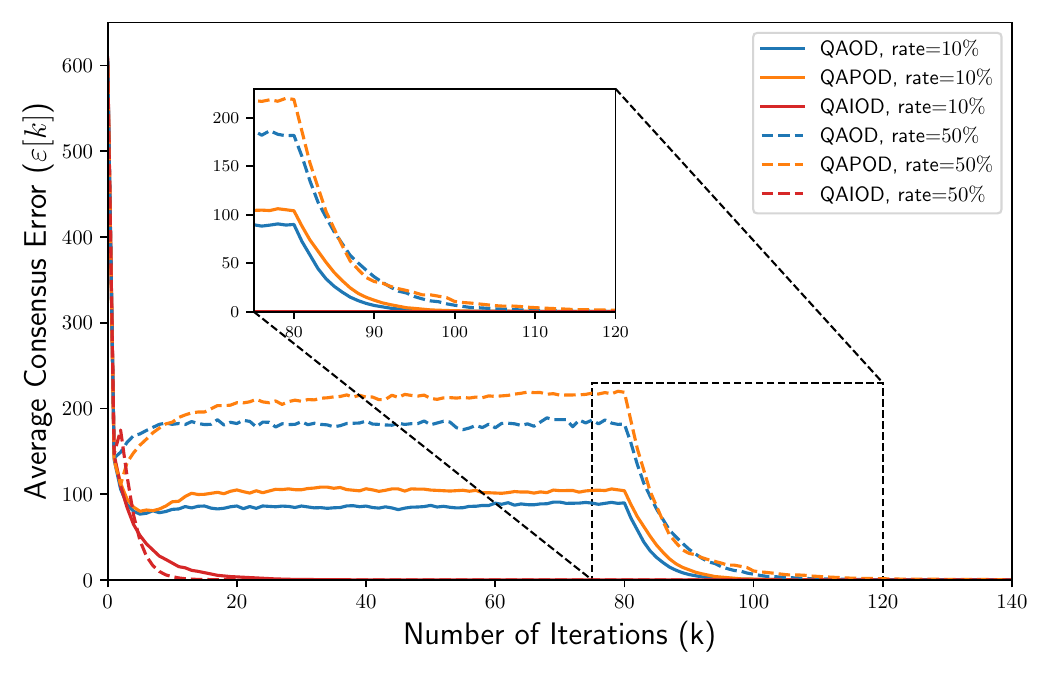} 
\caption{Evolution of error $\varepsilon[k]$ (shown in \eqref{error_plot}) over time $k$ averaged over $100$ random digraphs of $150$ potentially participating nodes for departing rates of $10\%$ and $50\%$, during the execution of QAOD, QAPOD, and QAIOD.} 
\label{fig:scenario3} 
\end{center} 
\end{figure} 

In Fig.~\ref{fig:scenario3}, during the operation of QAOD and QAPOD, for $k < 80$ the node states track the quantized average but do not stabilize due to continuous node arrivals and departures.  
Note that the value of $\varepsilon[k]$ for departure and arrival rates of $50\%$ is higher than that for $10\%$ since more nodes join and leave the network leading to greater perturbations in the quantized average of active nodes.  
For $k \geq 80$, however, the node states rapidly converge to the exact quantized average (i.e., $\varepsilon[k]$ becomes zero) for both departure and arrival rates of $10\%$ and $50\%$.  
Specifically, for a rate of $10\%$, both algorithms converge after approximately $100$ time steps, and for $50\%$ after approximately $120$ time steps.  
Finally, QAIOD achieves convergence after about $25$ time steps for a departure and arrival rates of $40\%$, and $40$ time steps for $10\%$.  
These results highlight its ability to maintain fast convergence even under high departure/arrival rates within a dynamic indefinitely OMAS. 

It is important to note here that, when the departure-handoff or joint-connectivity assumptions are violated, the mass-preservation and information-propagation properties used in the convergence analysis may not hold, and finite-time convergence to the exact quantized average is no longer guaranteed. 
We illustrate this behavior in Section~\ref{sec:stress_tests}.

\subsection{Stress Tests Under Assumption Violations}
\label{sec:stress_tests}

To further illustrate the applicability and limitations of QAOD, we provide stress tests below to demonstrate its behavior both when key assumptions are violated and under highly frequent membership changes.

\begin{figure}[t]
    \centering
    \includegraphics[width=8.4cm]{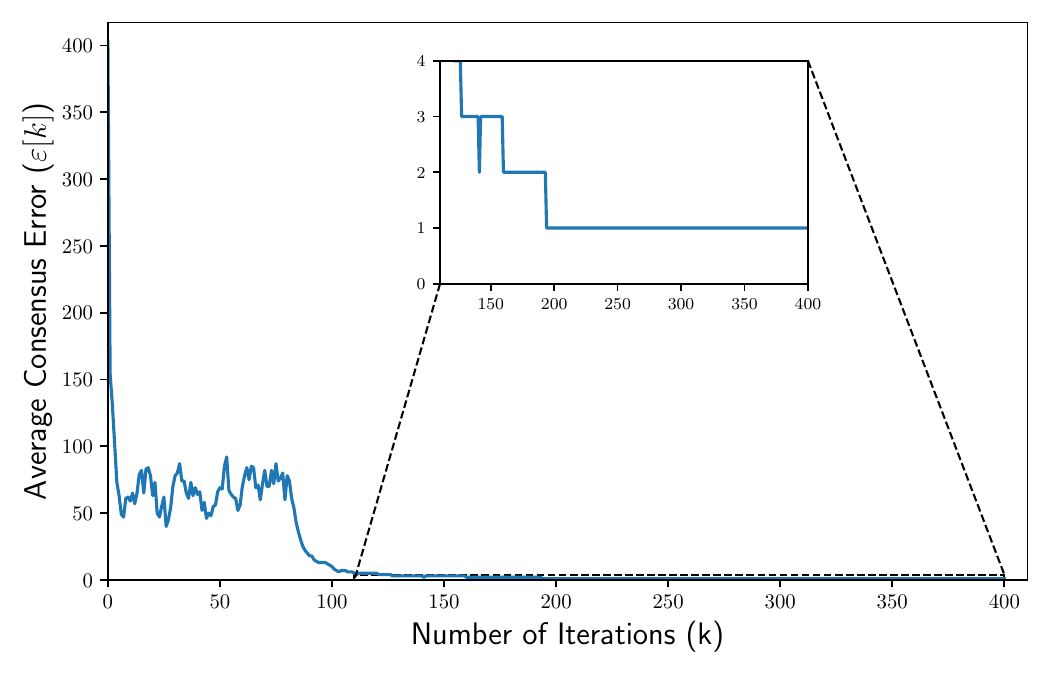}
    \caption{Evolution of the error $\varepsilon[k]$ over time $k$,
    averaged over $100$ random digraph realizations with $150$
    potentially participating nodes, during execution of QAOD when the handoff condition in Theorem~\ref{main_convergence_condition_theorem} is violated (i.e., some departing nodes have no remaining out-neighbor).}
    \label{handoff_violation_QAOD}
\end{figure}

\textit{Case--1: Violation of the departure-handoff condition.}
In Fig.~\ref{handoff_violation_QAOD}, we demonstrate QAOD in a setting in which all assumptions are satisfied, but some departing nodes have no remaining out-neighbor. 
For such a node, no handoff is possible and it leaves without transmitting its stored mass. 
As shown in Fig.~\ref{handoff_violation_QAOD}, the error $\varepsilon[k]$ does not converge to zero. 
This illustrates the necessity of the handoff condition mentioned in Theorem~\ref{main_convergence_condition_theorem} (i.e., when stored mass cannot be transferred before a node departs, exact mass preservation is lost). 

\begin{figure}[t]
    \centering
    \includegraphics[width=8.4cm]{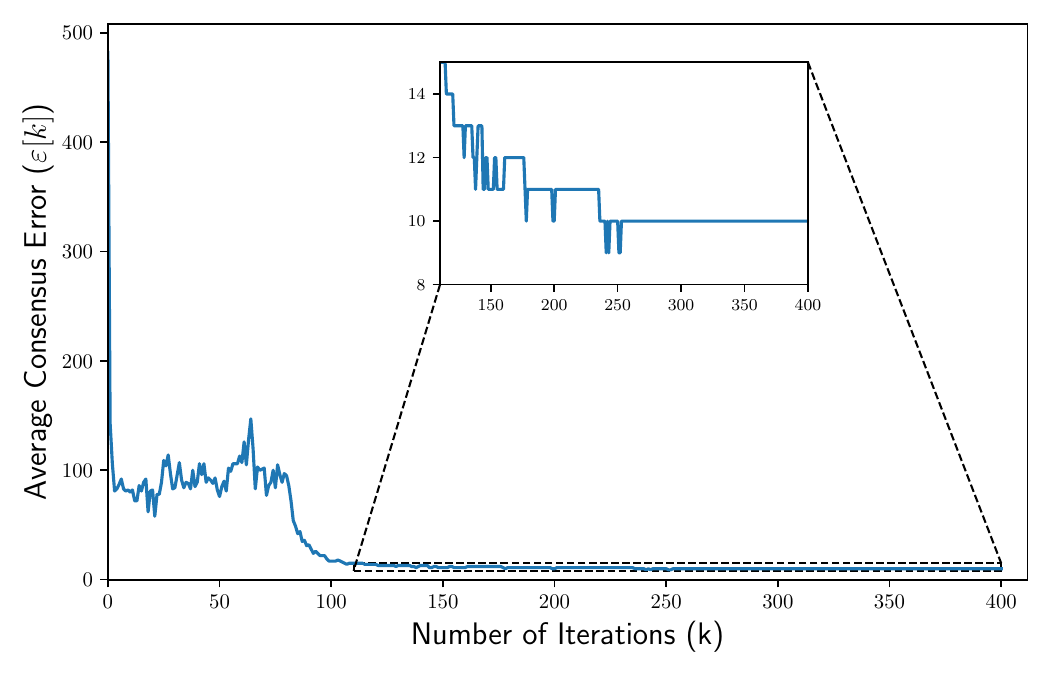}
    \caption{Evolution of the error $\varepsilon[k]$ over time $k$,
    averaged over $100$ random digraph realizations with $150$
    potentially participating nodes, during execution of QAOD when the post-stabilization joint-connectivity condition
    in Assumption~\ref{strong_connectivity_stable_union_graph} is violated.}
    \label{connectivity_violation_QAOD}
\end{figure}

\textit{Case--2: Violation of joint connectivity.}
In Fig.~\ref{connectivity_violation_QAOD}, we demonstrate QAOD with the departure-handoff condition satisfied at every departure, but with a post-stabilization communication sequence that does not satisfy Assumption~\ref{strong_connectivity_stable_union_graph}. 
As shown in Fig.~\ref{connectivity_violation_QAOD}, the error $\varepsilon[k]$ does not converge to zero. 
This illustrates that, without sufficient joint connectivity, the mass retained in the network cannot propagate to all active nodes, and the convergence guarantee of Theorem~\ref{main_convergence_condition_theorem} does not apply.

\begin{figure}[t]
    \centering
    \includegraphics[width=8.4cm]{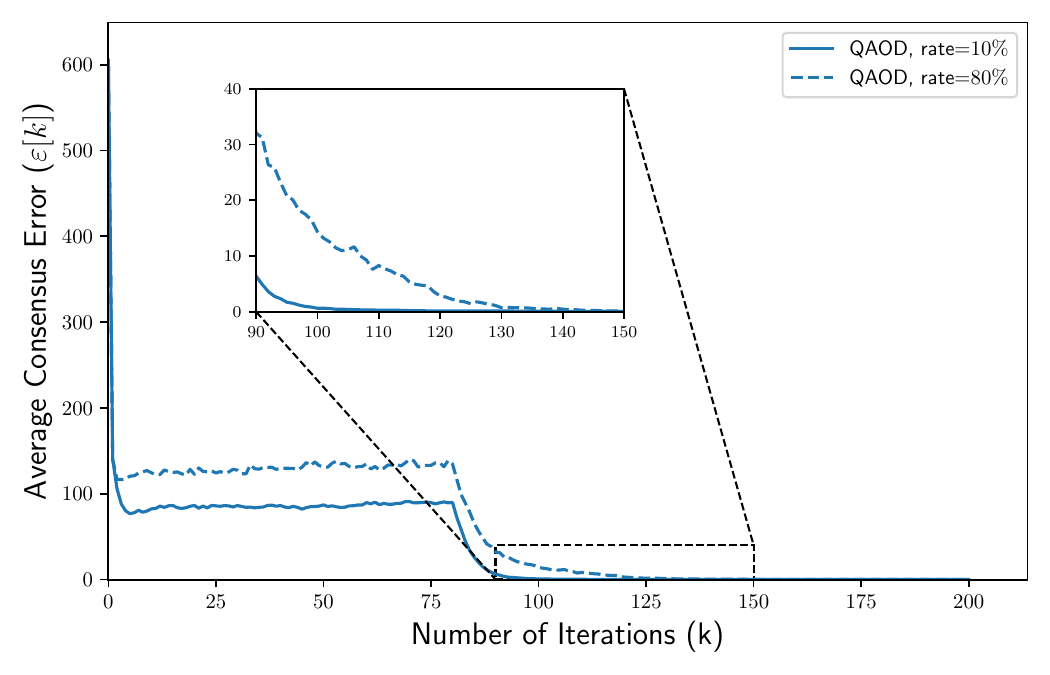}
    \caption{Evolution of the error $\varepsilon[k]$ over time $k$,
    averaged over $100$ random digraph realizations with $150$
    potentially participating nodes, during execution of QAOD for
    arrival and departure rates of $10\%$ and $80\%$ of the active-node
    set. All assumptions required by Theorem~\ref{main_convergence_condition_theorem} are satisfied.}
    \label{high_departure_rate_QAOD}
\end{figure}

\textit{Case--3: High arrival/departure rates.}
We demonstrate QAOD with arrival and departure rates equal to $80\%$ of the current active-node set at each time step and compare this case with the nominal rate of $10\%$. 
All assumptions required by Theorem~\ref{main_convergence_condition_theorem}, including the departure-handoff and joint-connectivity conditions, are satisfied in both cases. 
As shown in Fig.~\ref{high_departure_rate_QAOD}, QAOD converges in finite time for both rates. 
The $80\%$ case produces a larger transient error and a longer convergence time, but does not prevent convergence. 
This illustrates that QAOD can operate under highly frequent membership changes and exhibit finite-time convergence, provided that its assumptions remain satisfied. 

\subsection{Comparison with the Literature}

% \todo{note: this fig was done once, not 100 times averaged}

\begin{figure}[t] 
\begin{center}
\includegraphics[width=8.4cm]{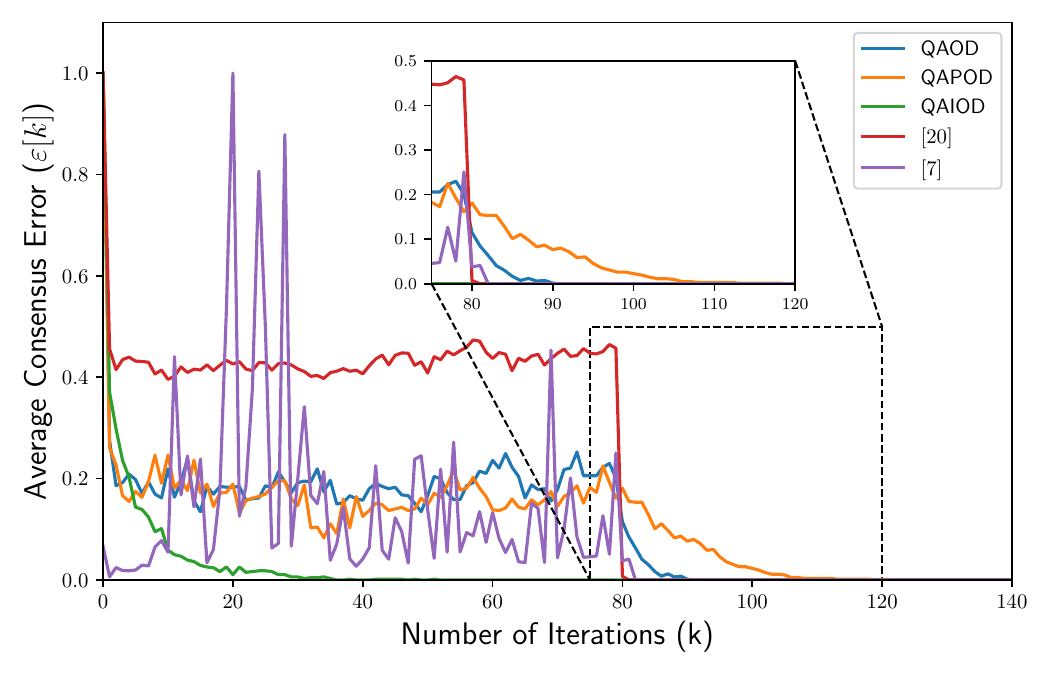} 
\caption{Evolution of normalized errors $\varepsilon[k]$, and $\varepsilon^\prime[k]$ over time $k$, for a random digraph of $150$ potentially participating nodes, during the execution of QAOD, QAPOD, QAIOD, \cite{2024:CDC_Themis_Open, 2024:CDC_Hadjic_Garcia}.} 
\label{fig:scenario4} 
\end{center} 
\end{figure}

We now compare QAOD, QAPOD, and QAIOD, against algorithms \cite{2024:CDC_Themis_Open, 2024:CDC_Hadjic_Garcia} from the literature. 
During the operation of QAOD, QAPOD, and QAIOD all parameters are consistent with \textit{Scenario~1} in Section~\ref{first_sec_results}, with QAPOD operating with $\bar{\tau}=5$. 
During the operation of \cite{2024:CDC_Themis_Open} and \cite{2024:CDC_Hadjic_Garcia} we consider the same parameters as in \textit{Scenario~1}, but the underlying digraph is static and strongly connected during all time steps. 
For our algorithms we plot the error $\varepsilon[k]$ defined in \eqref{error_plot}, setting $q_\text{alg}[k] = q[k]$ (see \eqref{goal}) for QAOD and QAPOD, and $q_\text{alg}[k] = q'[k]$ (see \eqref{eq:extend_goal}) for QAIOD. 
For \cite{2024:CDC_Themis_Open, 2024:CDC_Hadjic_Garcia} we plot the error $\varepsilon'[k]$ defined as $\varepsilon'[k]=||diag(\bm{\alpha}[k])\bm{z}[k]-\bm{1}q[k]||$ (from \cite[Section~V]{2024:CDC_Themis_Open}), where $\bm{\alpha}[k] \in \{0, 1\}^n$ is the indicator vector, where $\alpha_j[k] = 1$ if agent $v_j$ is active at time step $k$ (i.e., if $v_j \in \mathcal{V}[k]$) while $\alpha_j[k] = 0$, otherwise. 
\AR{Note that in Fig.~\ref{fig:scenario4}, both errors are normalized with respect to their initial values to allow for a direct and consistent comparison of their convergence trajectories (i.e.,  how quickly and effectively each approach eliminates disagreement across the active network).}

In Fig.~\ref{fig:scenario4}, we observe that QAOD, QAPOD, and QAIOD perform favorably compared to the real-valued algorithms in~\cite{2024:CDC_Themis_Open, 2024:CDC_Hadjic_Garcia}, which operate over open strongly connected networks. 
For time steps $k < 80$, all algorithms (except QAIOD) exhibit transient behavior, tracking the moving average without stabilizing due to ongoing node arrivals and departures. 
By design, QAIOD leverages Assumption~\ref{open_strong_connectivity_stable_union_graph} to converge to the quantized average of all historically active nodes after approximately $50$ time steps, demonstrating its robustness to continuous openness. 
After the network stabilizes at $k = 80$, algorithms QAOD, \cite{2024:CDC_Themis_Open}, and \cite{2024:CDC_Hadjic_Garcia} converge within $10$ time steps showcasing that QAOD exhibits comparable convergence speed as other real-valued algorithms. 
Furthermore, after $k = 80$, QAPOD converges in approximately $35$ time steps due to $\bar{\tau}=5$, illustrating the expected trade-off between computation delays and convergence speed. 
It is important to note, however, that while our algorithms may require more time steps to converge, they achieve it in finite time. 
In contrast, the real-valued algorithms from the literature only exhibit asymptotic convergence due to their real-valued operation. 
Furthermore, by exchanging quantized messages, our approach enables operation in bandwidth- and energy-limited environments, real-valued communication would be infeasible. 

% Crucially, while the algorithms from the literature converge asymptotically with real-valued communication, our proposed algorithms achieve finite-time convergence using only quantized message exchanges. This key feature makes them particularly well-suited for deployment in bandwidth- and energy-constrained environments.

% However, note here that despite the higher number of required time steps for convergence our algorithms converge in finite time (and the algorithms in the literature converge asymptotically due to their real-valued operation. 
% Additionally, our algorithms enable nodes to exchange quantized messages enabling them to operate over bandwidth limited and energy limited environments. 

\section{CONCLUSIONS}\label{future}

In this work we investigated the quantized averaging problem. 
We proposed three distributed algorithms: one for OMAS with dynamic communication links, one for OMAS with dynamic communication links and also node processing delays, and one for indefinitely OMAS with dynamic communication links. 
For all algorithms, we derived necessary and sufficient topological conditions and established finite‑time convergence under quantized communication. 
Finally, we demonstrated our algorithms' operation via an application to distributed sensor fusion for environmental monitoring, and showed that they compare favorably with existing algorithms. 

As future work we \AR{plan to conduct comprehensive quantitative communication-cost comparisons (including the number of bits per message, communication rounds, and total network load), together with a systematic sensitivity analysis of the convergence time with respect to the topology switching frequency (i.e., the joint-connectivity window length $T$). 
Moreover, we plan to develop finite-time quantized coordination algorithms for recency-weighted or sliding-window objectives over indefinitely open directed networks. 
Additionally, we will extend our algorithms to handle unreliable or delayed feedback channels (which would require an explicit acknowledgment/retransmission protocol, message identifiers, and a revised mass-preservation analysis).} 

% In this work, we investigated the distributed quantized averaging problem over open networks. 
% We proposed three algorithms. 
% Our first algorithm achieves quantized average consensus when the underlying open network features time-varying communication links. 
% Our second algorithm extends this setting to networks subject to node processing delays. 
% Our third algorithm operates over indefinitely open networks (i.e., persistent openness) with dynamic connectivity. 
% For all proposed algorithms, we derived necessary and sufficient topological conditions for convergence and proved their finite-time convergence under efficient quantized communication among nodes. 
% Our paper concludes with an application to distributed sensor fusion for target localization. 

% Future research will extend our frameworks to distributed optimization and learning over open networks with efficient quantized communication. 
% Promising directions also include resilient consensus and privacy-preserving strategies under intermittent node participation, for security-sensitive applications. 

\vspace{-0.3cm}

%%%%%%%%%%%%%%%%%%%%%%%%%%%%%%%%%%%%%%%%%%%%%%%%%%%%%%%%%
%
%
%   APPENDICES 
%
%
%%%%%%%%%%%%%%%%%%%%%%%%%%%%%%%%%%%%%%%%%%%%%%%%%%%%%%%%5

\section*{APPENDICES}

\appendix

% ===============================================
%
%
% Proof of ....
%
%
% ===============================================
\section{Proof of Theorem~\ref{main_convergence_condition_theorem}}
\label{convergence_Alg1_prob_P1}

Our proof is organized as follows.  
Initially, we will analyze the behavior of the underlying quantized averaging algorithm, adopting some notation from \cite[Theorem~1]{2022:Rikos_Hadj_Johan}.  
Next, we will examine the operation of QAOD separately for time steps \( k < k' \) and \( k \geq k' \) (where \( k' \) is defined in Assumption~\ref{existence_stable_time_step}).  
For \( k < k' \), we will establish how the sum of the initial states of active nodes is preserved despite departures and arrivals of potentially participating nodes.  
Finally, for \( k \geq k' \), we will demonstrate that QAOD enables active nodes to solve problem~\textbf{P1} in Section~\ref{sec:probForm}.

At any time step $k$, the operation of QAOD can be interpreted as the ``random walk'' of $n[k]$ ``tokens'' in a Markov chain with $n[k] = | \mathcal{V}[k] |$ states. 
This Markov chain is time-inhomogeneous, as its interconnections vary over time (and thus its transition probabilities also vary over time). 
Moreover, for time steps \( k < k' \) the number of states in the Markov chain also changes with time, while for \( k \geq k' \) the state set remains fixed (see Assumption~\ref{existence_stable_time_step}). 
Each active node $v_j$ at time step $k=0$ holds two ``tokens'': $T_j^{ins}$ (which is stationary) and $T_j^{out}$ (which performs a random walk).
At every time step $k$, they each contain a pair of values $y_j^{ins}[k]$, $z_j^{ins}[k]$, and $y_j^{out}[k]$, $z_j^{out}[k]$, respectively, for which it holds that $y_j^{ins}[0] = y_j^{out}[0] = x_j \in \mathbb{Z}$ and $z_j^{ins}[0] = z_j^{out}[0] = r_j = 1$. 
At each time step $k$, each node $v_j$ keeps the token $T_j^{ins}$ (i.e., it never transmits it) and transmits the token $T_j^{out}$, according to the nonzero probability $b_{lj}[k]$ it assigned to its outgoing edges $m_{lj}$. 
When $v_j$ receives one or more tokens $T_i^{out}$ from its in-neighbors, the values $y_i^{out}[k]$ and $y_j^{ins}[k]$ become equal (or differ by at most~$1$).
Then, $v_j$ transmits each received token $T_i^{out}$ to a randomly selected out-neighbor according to the nonzero probability $b_{lj}[k]$. 
Note here that during the operation of QAOD, for every time step $k$ we have 
\begin{equation}\label{sum_preserve_y}
\sum_{v_j \in \mathcal{V}[k]} (y^{out}_j[k] + y^{ins}_j[k]) = \sum_{v_j \in \mathcal{V}[k]} 2 x_j ,
\end{equation}
and 
\begin{equation}\label{sum_preserve_z}
\sum_{v_j \in \mathcal{V}[k]} (z^{out}_j[k] + z^{ins}_j[k]) = \sum_{v_j \in \mathcal{V}[k]} 2 r_j = 2 n[k] . 
\end{equation}

Next, we examine the impact of node arrivals and departures.  
We demonstrate that \eqref{sum_preserve_y} is maintained at every time step \( k < k' \), irrespective of changes in network membership due to nodes entering or leaving (the analysis for \eqref{sum_preserve_z} parallels that of \eqref{sum_preserve_y} and is omitted for brevity).  
It is important to note that our discussion is restricted to time steps \( k < k' \), since for \( k \geq k' \) the set of active nodes remains constant (see Assumption~\ref{existence_stable_time_step}) and no further arrivals or departures take place. 

\textbf{Arriving:} 
When a node \( v_j \) joins the network at time step \( k \) (where \( k < k' \)), it initializes its state and mass variables as specified by \eqref{init_variables}. 
In particular, \( v_j \) sets \( y_j[k+1] = 2x_j \) and \( z_j[k+1] = 2r_j \), and begins to interact with its neighbors in the following time step.
Each new node \( v_j \in \mathcal{A}[k] \) at time step \( k \) maintains two ``tokens'': \( T^{ins}_j \) (stationary) and \( T^{out}_j \) (which performs a random walk).
These tokens carry pairs of values: \( y^{ins}_j[k+1] \), \( z^{ins}_j[k+1] \) and \( y^{out}_j[k+1] \), \( z^{out}_j[k+1] \), for which it holds \( y^{ins}_j[k+1] = y^{out}_j[k+1] = x_j \in \mathbb{Z} \) and \( z^{ins}_j[k+1] = z^{out}_j[k+1] = r_j = 1 \).
Consequently, at time step \( k \), we have
\begin{equation}\label{eq:equation_arri}
    \sum\limits_{v_j \in \mathcal{A}[k]} y^{out}_j[k+1] + \sum\limits_{v_j \in \mathcal{A}[k]} y^{ins}_j[k+1] = 2 \sum\limits_{v_j \in \mathcal{A}[k]} x_j \ .
\end{equation}
Combining \eqref{eq:equation_arri} with \eqref{sum_preserve_y}, it follows that \eqref{sum_preserve_y} remains valid at time step \( k+1 \), ensuring that the sum preservation property holds even as new nodes enter the network.

\textbf{Departing:} 
Let us suppose that node $v_j$ departs from the network at time step $k$ and it has at least one remaining out-neighbor (i.e., $| \ \mathcal{N}^{+}_j[k] \cap \mathcal{R}[k] \ | \geq 1$). 
Node $v_j$ assigns nonzero probabilities $b_{lj}[k]$ as in \eqref{eq:depart_trans_prob}. 
It then computes the transmission variables as in \eqref{trans_vari}, and transmits them to one randomly chosen remaining out-neighbor. 
After this transmission, \( v_j \) leaves the network and becomes inactive in the next time step. 
It is important to note that if $v_j$ has no remaining out-neighbors (i.e., $| \ \mathcal{N}^{+}_j[k] \cap \mathcal{R}[k] \ | = 0$), it cannot compute $b_{lj}[k]$ and then transmit to a remaining out-neighbor. 
In this case, when $v_j$ leaves the network the information it holds will be lost. 
In such a case, valuable data about the initial states of other remaining nodes may be lost, potentially hindering the solution of problem~\textbf{P1}. 
Note now that if the tokens of the departing nodes are removed from the network (i.e., every departing node has at least one remaining out-neighbor) then we have  
\begin{equation}\label{eq:equation_depar}
\begin{aligned}
&\sum\limits_{v_j \in \mathcal{V}[k]} y^{out}_j[k] + \sum\limits_{v_j \in \mathcal{V}[k]} y^{ins}_j[k] \\ - \ & (\sum\limits_{v_j \in \mathcal{D}[k]} y^{out}_j[0] + \sum\limits_{v_j \in \mathcal{D}[k]} y^{ins}_j[0]) \\ = \ & 2(\sum\limits_{v_j \in \mathcal{V}[k]} x_j - \sum\limits_{v_j \in \mathcal{D}[k]} x_j), 
\end{aligned} 
\end{equation} 
for every $k < k'$, where $y^{ins}_j[0] = y^{out}_j[0] = x_j \in \mathbb{Z}$. 
Before leaving the network, each departing node $v_j \in \mathcal{D}[k]$ transmits its negative values of tokens of time step $k=0$ (i.e., $-(y^{out}_j[0] + y^{ins}_j[0])$, $-(z^{out}_j[0] + z^{ins}_j[0])$, where $-(y^{out}_j[0] + y^{ins}_j[0] ) = -2 x_j = -y_j[0]$, and $-(z^{out}_j[0] + z^{ins}_j[0]) = -2 r_j = -z_j[0]$) for eliminating the departing nodes' information. 
In addition, each departing node $v_j$ also transmits the values of tokens $T^{out}_j$, $T^{ins}_j$ it holds at the time step $k$, (i.e., transmits $y^{out}_j[k] + y^{ins}_j[k]$, and $z^{out}_j[k] + z^{ins}_j[k]$, where $y^{out}_j[k] + y^{ins}_j[k] = y_j[k]$, and $z^{out}_j[k] + z^{ins}_j[k] = z_j[k]$), for avoiding losing the other remaining nodes' information. 
Therefore, by transmitting the variables as defined in \eqref{trans_vari}, \eqref{eq:equation_depar} is satisfied. 
Moreover, \eqref{eq:equation_depar} is equivalent to 
\begin{equation}\label{eq:equation_depar_equa} 
    \sum\limits_{v_j \in \mathcal{R}[k]} y^{out}_j[k] + \sum\limits_{v_j \in \mathcal{R}[k]} y^{ins}_j[k] = 2 \sum\limits_{v_j \in  \mathcal{R}[k]} x_j , 
\end{equation}
for $k < k'$. 
Let us now assume for simplicity that no arrivals occur at time step $k$ (the case where arrivals and departures occur simultaneously at time step $k$ can be analyzed by combining equations \eqref{eq:equation_arri} and \eqref{eq:equation_depar_equa}). 
This means that \eqref{eq:equation_depar_equa} further yields
\begin{equation}\label{eq:equation_k+1}
\begin{aligned} 
&\sum\limits_{v_j \in \mathcal{V}[k+1]} y^{out}_j[k+1] + \sum\limits_{v_j \in \mathcal{V}[k+1]} y^{ins}_j[k+1] \\ & = 2\sum\limits_{v_j \in \mathcal{V}[k+1]} x_j ,  
\end{aligned}
\end{equation}
for every $k < k'$. 
As a result, \eqref{sum_preserve_y} continues to hold at time step \( k+1 \), ensuring the sum preservation property even when nodes depart from the network. 

% We now consider the network dynamics for time steps \( k \geq k' \). 
% By Assumption~\ref{existence_stable_time_step}, no additional node arrivals or departures occur beyond this point. 
% However, the communication links between nodes continue to change over time. 
% This means that our network $\mathcal{G}_d[k]$ is defined as \( \mathcal{G}_d[k] = (\mathcal{V}_{\mathcal{R}}, \mathcal{E}[k]) \) for \( k \geq k' \). 
% For \( \mathcal{G}_d[k] \), Assumption~\ref{strong_connectivity_stable_union_graph} is satisfied. 
% This means that the network is \( T \)-jointly strongly connected for all \( k \geq k' \). 
% Since \( \mathcal{G}_d[k] \) is \( T \)-jointly strongly connected and its node set remains unchanged for \( k \geq k' \), the network's convergence can be analyzed in the same fashion as in \cite[Theorem~1]{2022:Rikos_Hadj_Johan}. 
% The detailed proof is omitted for brevity, but the argument closely parallels that in the cited theorem.

\AR{We now consider the network dynamics for time steps \( k \geq k' \). 
By Assumption~\ref{existence_stable_time_step}, no additional node arrivals or departures occur beyond this point, and the set of active nodes stabilizes to \( \mathcal{V}_{\mathcal{R}} \). 
Hence, the network is given by \( \mathcal{G}_d[k] = (\mathcal{V}_{\mathcal{R}}, \mathcal{E}[k]) \) for all \( k \geq k' \). 
Under Assumption~\ref{strong_connectivity_stable_union_graph}, there exists an infinite sequence of recurrent time instants \( k' = t_0 < t_1 < t_2 < \dots \) with bounded interval lengths \( 1 \leq t_{m+1} - t_m \leq T \) such that the union graph over each interval satisfies \( \mathcal{G}_d^{t_m, \dots, t_{m+1}-1} = \mathcal{G}_d \), where \( \mathcal{G}_d = (\mathcal{V}_{\mathcal{R}}, \mathcal{E}^{\mathrm{nom}}) \) is a fixed strongly connected nominal digraph with diameter \( D^{\mathrm{un}} \). 
Since the node set \( \mathcal{V}_{\mathcal{R}} \) remains fixed for \( k \geq k' \) and the union graph equals the strongly connected nominal digraph \( \mathcal{G}_d \) over recurring intervals of length at most \( T \), information is guaranteed to propagate across any directed path in \( \mathcal{G}_d \) within at most \( D^{\mathrm{un}} T \) time steps. 
Therefore, the post-stabilization (i.e., $k \geq k'$) convergence of QAOD can be analyzed in the same fashion as in \cite[Theorem~1]{2022:Rikos_Hadj_Johan}. 
Combined with the global sum-preservation property maintained during all prior time steps \( k < k' \) (as established above), this guarantees that there exists a finite time step \( k_0 \geq k' \) such that every active node \( v_j \in \mathcal{V}_{\mathcal{R}} \) computes \( q^s_j[k] \in \{ \lfloor q[k] \rfloor, \lceil q[k] \rceil \} \) for all \( k \geq k_0 \).}

%%%%%%%%%%%%%%%%%%%%%%%%%%%%%%%%%%%%%%%%%%%%%%%%%%%%%%%%%%%%%%%%%%%%%%%%%%%%%%%
%%%%%%%%%%%%%%%%%%%%%%%%%%%%%%%%%%%%%%%%%%%%%%%%%%%%%%%%%%%%%%%%%%%%%%%%%%%%%%%

\section{Proof of Theorem~\ref{main_convergence_condition_theorem_process_delays}}
\label{convergence_Alg2_prob_P1_process}

The proof is similar to the proof of Theorem~\ref{main_convergence_condition_theorem}. 
For this reason, we only mention the differences in comparison to the proof of Theorem~\ref{main_convergence_condition_theorem}. 

Initially, let us note that the behavior of the underlying quantized averaging algorithm is similar to the one in QAOD. 
Regarding the operation of QAPOD, we will focus on time steps \( k < k' \) and \( k \geq k' \) (where \( k' \) is defined in Assumption~\ref{existence_stable_time_step}). 
we examine the impact of node arrivals and departures. 
We demonstrate that \eqref{sum_preserve_y} is maintained at every time step \( k < k' \), irrespective of changes in network membership due to nodes entering or leaving (similarly to the proof of Theorem~\ref{main_convergence_condition_theorem}, the analysis for \eqref{sum_preserve_z} parallels that of \eqref{sum_preserve_y} and is omitted for brevity). 
Similar to the proof of Theorem~\ref{main_convergence_condition_theorem}, our discussion will be restricted to time steps \( k < k' \) (since for \( k > k' \) the set of active nodes remains constant (see Assumption~\ref{existence_stable_time_step}) and no further arrivals or departures take place). 

Focusing on time steps \( k < k' \), we have the following analysis for the different operating modes. 

\textbf{Arriving:} 
The analysis here is similar to the analysis in Theorem~\ref{main_convergence_condition_theorem}. 

\textbf{Departing:} 
Let us suppose node $v_j$ departs from the network at time step $k$ and has at least one long-term remaining out-neighbor (i.e., $|\mathcal{N}^{+}_j[k] \cap \mathcal{R}'[k]| \geq 1$). 
Using the analysis outlined in the ``Departing'' case of Theorem~\ref{main_convergence_condition_theorem}, it follows that equation~\eqref{eq:equation_depar} holds for all $k < k'$, assuming $y^{ins}_j[0] = y^{out}_j[0] = x_j \in \mathbb{Z}$. 
When departing nodes execute the departing strategy (transmitting their stored tokens combined with tokens having values equal to the negative of their their initial states), equation~\eqref{eq:equation_depar} is equivalent to 
\begin{equation}\label{eq:equation_depar_equa_process_delays} 
    \sum\limits_{v_j \in \mathcal{R}'[k]} y^{out}_j[k] + \sum\limits_{v_j \in \mathcal{R}'[k]} y^{ins}_j[k] = 2 \sum\limits_{v_j \in  \mathcal{R}'[k]} x_j , 
\end{equation}
for $k < k'$. 
From~\eqref{eq:equation_depar_equa_process_delays}, we deduce that~\eqref{eq:equation_k+1} consequently holds. 
Therefore,~\eqref{sum_preserve_y} is preserved at time step $k+1$, ensuring the sum preservation property persists even in the presence of node departures.

\textbf{Departing Soon:} 
In this operating mode nodes inform their in-neighboring nodes for their status (i.e., that they are not in ``Long Term Remaining'' operating mode). 
So their in-neighbors do not transmit information to them. 
This means that the nodes that are in ``Departing Soon'' operating mode do not receive any information from their in-neighbors. 
Additionally, nodes that are in ``Departing Soon'' operating mode processes any stored information they received (from previous time steps). 
So they are not transmitting any information to their out-neighbors. 
In this case, we have that since nodes that are in ``Departing Soon'' operating mode do not transmit or receive information then we have that the sum preservation property in \eqref{sum_preserve_y} holds. 

\textbf{Long-Term Remaining:}
The operation of long-term remaining nodes is similar to the operation of the underlying quantized averaging algorithm (also presented in \cite[Theorem~1]{2022:Rikos_Hadj_Johan}). 
In particular, long-term remaining nodes process the information received from their in-neighbors and, within at most $\overline{\tau}$ time steps, transmit their stored information to other long-term remaining nodes or themselves. 
As a result, equation~\eqref{sum_preserve_y} is satisfied by all long-term remaining nodes at every time step $k$.

We now analyze the network dynamics for time steps \( k \geq k' \). 
According to Assumption~\ref{existence_stable_time_step}, there are no further node arrivals or departures after this time. 
This means that for every active node $v_j$, it holds that $\rho_j^l[k] \gg \bar{\tau}_{j}$ for all time steps \( k \geq k' \). 
Consequently, all active nodes become long-term remaining nodes and operate by following the corresponding strategy. 
Despite this, the communication links between nodes continue to change over time. 
Under Assumption~\ref{strong_connectivity_stable_union_graph}, the network remains \( T \)-jointly strongly connected for all \( k \geq k' \). 
Since \( \mathcal{G}_d[k] \) maintains \( T \)-joint strong connectivity and a fixed node set for \( k \geq k' \), the convergence analysis shares similarities with the one in \cite[Theorem~1]{2022:Rikos_Hadj_Johan}. 
However, note here that the key difference from the analysis in \cite[Theorem~1]{2022:Rikos_Hadj_Johan} is that both long-term remaining nodes and departing nodes perform transmissions after at most $\bar{\tau}$ time steps (see~\eqref{upper_bound_processingdelay_opendynamnamnam}), rather than at every time step. 
As a result, accounting that transmissions among active nodes occur every $\bar{\tau}$ time steps, we have that the convergence analysis follows the same structure as in \cite[Theorem~1]{2022:Rikos_Hadj_Johan} and in \cite[Theorem~4]{rikos2024distributed}.

\section{Proof of Theorem~\ref{main_convergence_condition_theorem_alwaysopen}}
\label{convergence_Alg3_prob_P2}

% During the operation of \AR{QAIOD} we have that $n = |\mathcal{V}^\prime|$ (where $\mathcal{V}^\prime = \{v_1, v_2, . . ., v_n\}$) potentially participating nodes are following the Initialization procedure of \AR{QAIOD}. 
% Let us note now that during the operation of \AR{QAIOD}, we have that the set of potentially participating nodes $\mathcal{V}^\prime$ is split into 4 subsets. 
% Specifically, $\mathcal{V}^\prime$ is split into (i) $\mathcal{X}_1$ nodes that never become active during the operation of the algorithm, (ii) $\mathcal{X}_2$ nodes that become active for a finite number of time steps in total and after a specific time step they remain inactive and never become active again, (iii) $\mathcal{X}_3$ nodes that become active frequently, they may become inactive for a finite number of time steps, but after a finite number of steps they always become active again, and (iv) $\mathcal{X}_4$ nodes that remain active always and never become inactive. 
During the operation of QAIOD, we consider $n = |\mathcal{V}'|$ potentially participating nodes with $\mathcal{V}' = \{v_1,\ldots,v_n\}$, all of which follow the Initialization procedure of QAIOD. 
Throughout the operation of the algorithm, the set $\mathcal{V}'$ is partitioned into four subsets: 
(i) $\mathcal{X}_1$, nodes that never become active;
(ii) $\mathcal{X}_2$, nodes that are active for only a finite number of time steps and thereafter remain permanently inactive;
(iii) $\mathcal{X}_3$, nodes that become active infinitely often, possibly interspersed with finitely many inactive intervals, but ultimately reactivating;
and (iv) $\mathcal{X}_4$, nodes that remain active at all times and never become inactive. 
The sets are defined as $\mathcal{X}_1 = \bigl\{\, v_j \in \mathcal{V}' \ | \ v_j \notin \mathcal{V}[k] \ \text{for all} \ k \in \mathbb{N} \,\bigr\}$, $\mathcal{X}_2 = \bigl\{\, v_j \in \mathcal{V}' \;\big|\; \exists\, \iota_j \in \mathbb{N}\ \text{s.t.}\ v_j \in \mathcal{V}[k]\ \text{for some}\ k \le \iota,\ \text{and}\ v_j \notin \mathcal{V}[k]\ \text{for all}\ k > \iota_j \,\bigr\}$, $\mathcal{X}_3 = \bigl\{\, v_j \in \mathcal{V}' \;\big|\; v_j \in \mathcal{V}[k]\ \text{for infinitely many}\ k,\ \text{and}\ v_j \notin \mathcal{V}[k]\ \text{for at most finitely many}\ k \,\bigr\}$, and $\mathcal{X}_4 = \bigl\{\, v_j \in \mathcal{V}' \;\big|\; v_j \in \mathcal{V}[k]\ \text{for all}\ k \in \mathbb{N} \,\bigr\}$. 
From the definition of the sets, we have that $\mathcal{V}^\prime = \mathcal{X}_1 \cup \mathcal{X}_2 \cup \mathcal{X}_3 \cup \mathcal{X}_4$. 

Initially, let us note that the behavior of the underlying quantized averaging algorithm is similar to the one in QAOD. 
Let us now define $\iota'$ as 
\begin{equation}\label{max_stop_participate}
    \iota' = \max_{v_j \in \mathcal{X}_2} \iota_j, 
\end{equation}
denoting the last time step at which any node in $\mathcal{X}_2$ is active (i.e., after time step $\iota'$ all nodes in $\mathcal{X}_2$ remain permanently inactive). 
We will analyze the operation of QAIOD for time steps $k \leq \iota'$ and $k > \iota'$. 

For time steps $k \leq \iota'$, during the operation of QAIOD each node $v_j \in \mathcal{X}_2 \cup \mathcal{X}_3 \cup \mathcal{X}_4$ is following one of the three operating modes: ``Arriving'', ``Departing'', or ``Remaining''. 
Following an analysis similar to Theorem~\ref{main_convergence_condition_theorem}, we can show that the following relations 
\begin{equation}\label{sum_preserve_y_for_alg3}
\sum_{v_j \in \mathcal{V}[k]} (y^{out}_j[k] + y^{ins}_j[k]) = \sum_{v_j \in \mathcal{H}[k]} 2 x_j ,
\end{equation}
and 
\begin{equation}\label{sum_preserve_z_for_alg3}
\sum_{v_j \in \mathcal{V}[k]} (z^{out}_j[k] + z^{ins}_j[k]) = \sum_{v_j \in \mathcal{H}[k]} 2 r_j . 
\end{equation}
where $\mathcal{H}[k]$ is defined in \eqref{set_active_until_k}, hold during every time step $k \leq \iota'$ irrespective of changes in network membership due to nodes entering or leaving. 
This means that the number of tokens of all nodes in the set $\mathcal{L} \subseteq \mathcal{X}_2 \cup \mathcal{X}_3 \cup \mathcal{X}_4$ is maintained despite arrivals and/or departures of nodes in the set $\mathcal{X}_2 \cup \mathcal{X}_3$.

For all time steps $k > \iota'$, under QAIOD each node $v_j \in \mathcal{X}_3 \cup \mathcal{X}_4$ operates in exactly one of the three modes: ``Arriving'', ``Departing'', or ``Remaining''. 
Following a similar analysis as Theorem~\ref{main_convergence_condition_theorem}, we have that \eqref{sum_preserve_y_for_alg3} and \eqref{sum_preserve_z_for_alg3} continue to hold. 
\AR{Since $|\mathcal{V}^\prime| = n < \infty$, let $\iota^{\prime\prime}$ be the finite time step after which no new node $v_j \in \mathcal{X}_3 \cup \mathcal{X}_4$ enters the network for the first time.} 
Let $(\gamma - 1) \in \mathbb{N}$ denote the maximum number of consecutive time steps during which any node $v_j \in \mathcal{X}_3 \cup \mathcal{X}_4$ may remain inactive. 
Then, for every time step $k > \iota^{\prime\prime}$, each node $v_j \in \mathcal{X}_3 \cup \mathcal{X}_4$ becomes active at least once within the interval $\{k, \ldots, k+\gamma\}$. 
Consequently, for every $k > \iota^{\prime\prime}$, we have that 
\begin{equation}\label{all_active_set_P2}
    \mathcal{I}^{\gamma}[k] = \mathcal{X}_3 \cup \mathcal{X}_4 ,
\end{equation}
where $\mathcal{I}^{\gamma}[k]$ is defined in \eqref{subset_in_L_nodes}. 
Moreover, by Assumption~\ref{open_strong_connectivity_stable_union_graph}, there exists $\gamma' \in \mathbb{N}$ with $\gamma' \geq \gamma$ such that, over the interval $\{k, \ldots, k+\gamma'\}$ (for $k > \iota^{\prime\prime}$), every edge between nodes that were active at least once in that interval appears at least once. 
Using \eqref{all_active_set_P2} and Assumption~\ref{open_strong_connectivity_stable_union_graph}, it follows that for $k > \iota^{\prime\prime}$, the \textit{full virtual union digraph} defined as 
\begin{equation}\label{full_virtual_union_digraph_P2}
\widehat{\mathcal{G}}'_d[k] = \big(\mathcal{I}^{\gamma}[k], \mathcal{Q}^{\gamma'}[k]\big) ,
\end{equation}
with $\mathcal{Q}^{\gamma'}[k]$ defined in \eqref{subset_in_L_edges}, coincides with the nominal digraph $\widehat{\mathcal{G}}'_d = \big(\mathcal{X}_3 \cup \mathcal{X}_4, \mathcal{Q}^{\gamma'}[k]\big)$, namely the digraph containing all potentially active nodes (all nodes in $\mathcal{X}_3 \cup \mathcal{X}_4$) and all possible edges between them. 
Let us also assume that $\widehat{\mathcal{G}}'_d$ is strongly connected. 
For every $k > \iota^{\prime\prime}$, considering the strongly connected digraph $\widehat{\mathcal{G}}'_d[k]$ in \eqref{full_virtual_union_digraph_P2}, the proof proceeds as in \cite[Theorem~1]{2022:Rikos_Hadj_Johan}, with the difference being that we use $\widehat{\mathcal{G}}'_d[k]$ to compute the probability that a token is at node $v_i \in \mathcal{X}_3 \cup \mathcal{X}_4$ after $\gamma'$ steps. 
The detailed proof is omitted for brevity, as the remaining of the proof closely parallels the cited theorem.

% ------------------------------------------------------------------------------
% Bibliography
% ------------------------------------------------------------------------------

\bibliographystyle{plain}        % Include this if you use bibtex 

\bibliography{reference}           % and a bib file to produce the 

\end{document}